\documentclass[aps,prd,11pt,preprint,nofootinbib,showkeys]{revtex4-2}
\usepackage{amsmath,amssymb,amsthm,mathrsfs,bbm}
\usepackage{latexsym,amscd,amsbsy,amsfonts,dsfont}
\usepackage{graphicx}
\usepackage[utf8]{inputenc}
\usepackage{subfigure}  
\usepackage{float}
\usepackage{slashed}
\usepackage{cancel}
\usepackage{multirow}
\usepackage{soul}
\usepackage{physics}
\usepackage{comment}
\usepackage[usenames,dvipsnames]{xcolor}
\usepackage[
      colorlinks=true,
      linktocpage=true,
      breaklinks=true,
      linkcolor=Aquamarine,
      urlcolor=Purple,
      filecolor=black,
      citecolor=Purple,
      menucolor=black,
      pdfstartview=FitV,
      bookmarksopen=true
      ]{hyperref}

\newcommand{\jordanf}{\text{(J)}}
\newcommand{\einsteinf}{\text{(E)}}

\makeindex

\begin{document}

\title{Unimodular Gravity vs General Relativity: A status report }

\author{Ra\'ul Carballo-Rubio}
\email{raul@sdu.dk}
\affiliation{CP3-Origins, University of Southern Denmark, Campusvej 55, DK-5230 Odense M, Denmark}
\affiliation{Florida Space Institute, University of Central Florida, 12354 Research Parkway, Partnership 1, 32826 Orlando, FL, USA}
\author{Luis J. Garay}
\email{luisj.garay@ucm.es}
\affiliation{Departamento de F\'{\i}sica Te\'orica and IPARCOS, Universidad Complutense de Madrid, 28040 Madrid, Spain}
\author{Gerardo Garc\'ia-Moreno}
\email{ggarcia@iaa.es}
\affiliation{Instituto de Astrof\'{\i}sica de Andaluc\'{\i}a (IAA-CSIC), Glorieta de la Astronom\'{\i}a, 18008 Granada, Spain}

\begin{abstract}
Unimodular Gravity (UG) is an alternative to General Relativity (GR) which, however, is so closely related to the latter that one can wonder to what extent they are different. The different behavior of the cosmological constant in the semiclassical regimes of both frameworks suggests the possible existence of additional contrasting features. UG and GR are based on two different gauge symmetries: UG is based on transverse diffeomorphisms and Weyl rescalings (WTDiff transformations), whereas GR is based on the full group of diffeomorphisms. This difference is related to the existence of a fiduciary background structure, a fixed volume form, in UG theories. In this work we present an overview as complete as possible of situations and regimes in which one might suspect that some differences between these two theories might arise. This overview contains analyses in the classical, semiclassical, and quantum regimes. When a particular situation is well known we make just a brief description of its status. For situations less analyzed in the literature we provide here more complete analyses. Whereas some of these analyses are sparse through the literature, many of them are new. Apart from the completely different treatment they provide for the cosmological constant problem, our results uncover no further differences between them. We conclude that, to the extent that the technical naturalness of the cosmological constant is regarded as a fundamental open issue in modern physics, UG is preferred over GR since the cosmological constant is technically natural in the former.
\end{abstract}


\maketitle
 
\tableofcontents

\section{Introduction}
\label{Sec:Introduction}

The first work on Unimodular Gravity (UG) dates back to Einstein, who wrote its traceless equations for a metric with fixed determinant (in suitable coordinates) $\abs{g} = 1$ for the first time in~\cite{Einstein1919}, with the first translation of this article appearing in~\cite{Einstein1952}. Einstein pointed out the benefit of having the cosmological constant as an integration constant. However, the condition $\abs{g} = 1$ reduces the gauge group to only the subset of transverse diffeomorphisms (those preserving the volume form associated with the metric determinant) contained within the set of all possible diffeomorphisms. Furthermore, as it was pointed out in~\cite{Unruh1989}, it is possible to consider a direct generalization of this theory to make everything coordinate independent, by introducing a fixed background volume form and working with an unconstrained metric. This makes the theory invariant under Weyl rescalings of the unconstrained metric $g_{\mu \nu}$. Actually, it makes just the conformal structure of the metric fluctuate, as in UG. We will refer as WTDiff-invariant theories or UG theories to theories in which this volume form is written explicitly and display this combination of gauge symmetries: Weyl rescalings of the metric and transverse diffeomorphisms. 

Unimodular gravity has been studied in different contexts but, to our knowledge, there is not a source in which a systematic comparative of this WTDiff gauge invariance principle with a Diff gauge invariance principle has been presented. In UG analysis, special interest has been devoted to the differential behaviour of the cosmological constant in both theories~\cite{vanderBij1981,Zee1983,Wilczek1983,Weinberg1988,Smolin2009,Alvarez2010,Barcelo2014,Carballorubio2015}. The main difference found is that the observed value for the cosmological constant is technically natural, in a sense that we will discuss extensively in this work, for UG. For Diff-invariant theories of gravity coupled to the Standard Model of particle physics, the cosmological constant is not technically natural due to the huge hierarchy of scales between the cosmological constant and the Planck scale (or even the electroweak scale). However, it is an open question whether these two symmetry principles lead to further differences beyond this one.

In this work we systematically address the question of the equivalence between Diff-invariant theories and WTDiff-invariant theories. For that purpose, we have looked at several regimes and situations in which both theories might look differently, with the aim of being as exhaustive as possible. Whereas some of these regimes and situations were already studied in the literature, we have found convenient to critically review them here. Although will refer the reader to the original literature in those sections for technical details, we will discuss them within the unifying global perspective that we present here. Other sections include completely new material and we give plenty of details on the computations and more extensive discussions on the results. We find convenient to include everything together in a single source in order to fill a gap present in the literature.

Here is a brief outline of the remain of the work. Section~\ref{Sec:Classical_Theory} focuses on the classical aspects of both theories. We begin discussing linear theories for spin-2 fields in Subsec.~\ref{Subsec:Classica_Linear_Theory}, with special emphasis on the theories with the maximum amount of gauge symmetry possible: either the linearized versions of WTDiff and Diff-invariant theories. We then move on to their simplest non-linear completions in Subsec.~\ref{Subsec:Einstein_Hilbert}, either General Relativity or its closest version in WTDiff. The natural extensions to higher-derivative theories obtained from higher curvature terms in the Lagrangian are analyzed in detail in Subsec.~\ref{Subsec:Higher_Derivatives}. Subsec.~\ref{Subsec:Bootstrapping} is devoted to the study of the bootstrapping of these linear spin-2 field theories to obtain self-consistent non-linear extensions for both kind of theories. In Subsec.~\ref{Subsec:Nonminimal_couplings} we analyze the role of non-minimal couplings and the choice of frames on both theories. Finally, in Subsec.~\ref{Subsec:Non_Conserv} we give a discussion on how the theories might differ once one allows for a non-conserved energy-momentum tensor in UG. That concludes our discussion of classical aspects of both theories. Section~\ref{Sec:Semiclassical} is devoted to the study of semiclassical effects on both theories. We compare their behaviour with respect to the cosmological constant in Subsec.~\ref{Subsec:CC} and Sakharov's induced gravity program in Subsec.~\ref{Subsec:Sakharov}. In Section~\ref{Sec:Perturbative_QFT} we move on to the quantum realm and the perturbative quantization of the theories. In Subsec.~\ref{Subsec:Quantization} we deal with the perturbative quantization of both theories on top of flat spacetime; in Subsec.~\ref{Subsec:Constructibility} we study graviton scattering amplitudes of both theories as well as their constructibility via recursion relations, and we conclude in Subsec.~\ref{Subsec:Asymptotic_Safety} with a brief overview of the asymptotic safety program applied to UG. Section~\ref{Sec:Strings} discusses the embedding of the theory in string theory from the point of view of the graviton scattering amplitudes. The embedding through the analysis of cancellation of conformal anomalies ($\beta$-functionals computations) is much more convoluted and we will report it elsewhere, in order to keep the extension of this article manageable. We analyze then the Euclidean path integral approach to a non-perturbative quantization in Section~\ref{Sec:Non_Perturbative_Path_Integral}. We review in Subsec.~\ref{Subsec:Conformal_Factor} the conformal factor problem appearing in Diff-invariant. In Subsec.~\ref{Subsec:Longitudinal_Diff} we show how the problem disappears in WTDiff-invariant theories, whereas a new problem due to the absence of longitudinal diffeomorphisms as gauge symmetries appears. We summarize in Section~\ref{Sec:Conclusions} and draw up the conclusions that can be taken from this work. We have also added an Appendix~\ref{App:UG_Formulations} including a short review of alternative formulations of UG and how they are equivalent to the unconstrained formalism that we pursue here.

\emph{Notation and conventions:} Our convention for the signature of the metric is $(-,+,...,+)$. Tensor objects will be represented by bold symbols, whereas their components in a given basis will be written with the same (not bold) symbol and indices, e.g., the Minkowski metric $\boldsymbol{\eta}$ will be represented in components as $\eta_{\mu \nu}$. At some point, we will introduce an auxiliary constrained metric $\boldsymbol{\tilde{g}}$ written in terms of the unconstrained metric $\boldsymbol{g}$. The covariant derivatives compatible with those metrics are denoted as $\tilde{\nabla}$ and $\nabla$, respectively. Their determinants will be denoted as $g$ and $\tilde{g}$, respectively. Curvature quantities like the Riemann tensor are defined following Misner-Thorne-Wheeler's conventions~\cite{Misner1974} and we will specify explicitly the metric it depends on, e.g. $R^{\alpha}_{\ \beta  \gamma \delta} (\boldsymbol{g})$. The gravitational Newton constant $\kappa^2/(8\pi)$ in $D+1$ dimensions and the Planck mass $M_\textsc{p}$ are related via the expression $\kappa^2  = M_\textsc{p}^{-(D-1)}$. In order to discuss the embedding of a WTDiff principle in string theory, we need to work in arbitrary spacetime dimension, $D+1$. Hence, in the parts of Section~\ref{Sec:Classical_Theory} needed for the discussion of Section~\ref{Sec:Strings} we work in arbitrary spacetime dimension $D+1$ for convenience. In the rest of the sections, we will restrict to $D = 3$.

\section{Classical theory}
\label{Sec:Classical_Theory}

\subsection{Linear spin-2 theories LWTDiff and LDiff}
\label{Subsec:Classica_Linear_Theory}

Elementary particles are associated with unitary irreducible representations of the proper orthocronous Poincar\'e group. Such representations can be classified according to the mass $m$ of the particle and the quantum numbers of its little group~\cite{Wigner1939,Weinberg1995,Bekaert2006}. For massive particles \mbox{$m \neq 0$}, we can always choose a reference frame such that they are at rest and their momentum reads $P^{\mu} = (m, 0 , ..., 0)$. The little group corresponds to the transformations that leave invariant this choice of momentum: the group $SO(D)$ of $D$-dimensional rotations. For the special case $D=3$, the label coming from the little group is related to the angular momentum $j$, cataloguing the projective representations of $SO(3)$. For a particle with angular momentum $j$, we have $2j +1$ states labelled by the polarization $s$, which runs from $s = -j$ to $s = +j$ in steps of one unit. For massless particles, $m=0$, the best we can do is to choose a reference frame such that the momentum reads $P^{\mu} = (E, 0 , ... , 0 ,E)$, being $E$ the energy of the particle. Now, the little group corresponds to the group $ISO(D-1)$, i.e., the $(D-1)$-dimensional Euclidean group. Again, for the special case $D=3$, the corresponding representations are labelled again with the angular momentum $j$ but just the states with polarizations $s = \pm j$ survive. Thus, a massless particle with angular momentum $j = 2$, which is what we will understand as a graviton in $D=3$ in the following, will have just two physical polarizations.

In a particle physics approach to gravity, being gravity just an interacting field theory of gravitons propagating on top of flat spacetime, the graviton field, represented as $h^{\mu \nu}$, is the mediator of the gravitational interaction. The first step in writing down such theory is to define a representation for these spin-2 fields. The naive way in which we would do it is defining an object which just contains the two physical polarizations of the graviton. From the operational point of view, we would like such an object to be a tensor in order for it to transform simply under coordinate transformations. However it is impossible to keep an object carrying just these two polarizations as a tensor, as the Lorentz transformations do not leave this reduced space invariant, for details see Section II in~\cite{Barcelo2014}.

Thus, we are left with two options in order to bypass the problem: treat $h^{\mu \nu}$ as a non-tensorial object or preserve such character but consider the whole space of symmetric tensors to be redundant, i.e., declare that configurations related by a concrete (gauge) transformation represent the same physical configuration. This second option is the one typically chosen in practice for its operational simplicity.  

To further complicate the matter, the object that naively would describe a graviton field, the two-index symmetric tensor $h^{\mu \nu}$, does not carry an irreducible representation of the Lorentz group. The $\frac{(D+1)(D+2)}{2}$-dimensional space it spans can be decomposed into irreducible representations of the Poincar\'e group as $\frac{(D+1)(D+2)}{2} = \frac{1}{2}(D+2)(D-1) \oplus D \oplus 1 \oplus 1$. To ensure that just the two physical polarizations of the field are taken into account, we need to introduce gauge symmetries and constraints on the field $h^{\mu \nu}$. It is always possible to relax the constraints on the field $h^{\mu \nu}$ at the expense of enlarging the set of gauge transformations. This is the procedure usually taken when writing down field theories, like quantum electrodynamics, for instance. Hence, we will restrict ourselves to discuss the representations of $h^{\mu \nu}$ carrying the maximal amount of gauge symmetries available. This problem was studied in detail in~\cite{Alvarez2006}. The most general action that one can write, up to boundary terms, for a massless spin-2 field $h^{\mu \nu}$ is
\begin{align}
    S = \int d^{D+1}x \sqrt{- \eta} \sum_{n=1}^{4} M^{\alpha \tau}_{(n)   \beta  \gamma \rho \sigma}  \nabla^{(\eta)}_{\alpha} h^{\beta \gamma} \nabla^{(\eta)}_{\tau} h^{\rho \sigma},
    \label{masslessspin2}
\end{align}
where $\nabla^{(\eta)}$ represents the covariant derivative compatible with the flat spacetime metric $\boldsymbol{\eta}$ and the tensors $M^{\alpha \tau}_{(n) \beta \gamma \rho \sigma}$ depend on three real parameters $\xi_1, \xi_2, \xi_3 \in \mathbb{R}$ and can be expressed as
\begin{align}
    & M^{\alpha \tau}_{(1)  \beta \gamma \rho \sigma} = - \frac{1}{4} \eta^{\alpha \tau} \eta_{\beta (\rho} \eta_{\sigma) \gamma}, \nonumber \\
    & M^{\alpha \tau}_{(2)   \beta \gamma \rho \sigma} =  \frac{1}{4} ( 1 + \xi_1) \left( \eta_{\beta (\rho} \delta^{\alpha}_{\sigma)} \delta^{\tau}_{\ \gamma} + \eta_{\gamma  (\rho} \delta^{\alpha}_{\sigma)} \delta^{\tau}_{\ \beta } \right), \nonumber \\
    & M^{\alpha \tau}_{(3)   \beta \gamma \rho \sigma} = - \frac{1}{4}( 1 + \xi_2) \left( \delta^{\alpha}_{\ (\beta} \delta^{\tau}_{\ \gamma)} \eta_{\rho \sigma} +  \delta^{\tau}_{\ (\rho} \delta^{\alpha}_{\ \sigma)} \eta_{\beta \gamma}  \right), \nonumber \\
    & M^{\alpha \tau}_{(4)   \beta \gamma \rho \sigma} = \frac{1}{4} (1 + \xi_3) \eta^{\alpha \tau} \eta_{\beta \gamma} \eta_{\rho \sigma }.
    \label{Eq:Tensors}
\end{align}
The maximum amount of gauge symmetry is found just for two sets of parameters~\cite{Alvarez2006}. On the one hand, for $\xi_1 = \xi_2 = \xi_3 = 0$, we recover Fierz-Pauli theory~\cite{Fierz1939}. As it is well known, such theory is invariant under linearized diffeomorphisms generated by arbitrary vector fields $\chi^{\mu}$ as 
\begin{equation}
    h^{\mu \nu}_{\xi} = h^{\mu \nu} + \eta^{\mu \tau} \nabla^{(\eta)}_{\tau} \chi^{\nu} +  \eta^{ \nu \tau} \nabla^{(\eta)}_{\tau} \chi^{\mu}.
\end{equation}
On the other hand, for $\xi_1 = 0, \ \xi_2 = -1 /2, \  \xi_3 = -5/8$, we have the so-called WTDiff theory. It is invariant under local Weyl scalings and transverse diffeomorphisms. They are generated by arbitrary scalar fields $\varphi$ and transverse vector fields $\chi^{\mu}$, i.e. those obeying $ \nabla^{(\eta)}_{\rho} \chi^{\rho} = 0$, as  
\begin{equation}
    h^{\mu \nu}_{( \varphi,\xi )} = h^{\mu \nu} + \frac{1}{2} \eta^{\mu \nu} \varphi + \eta^{\mu \tau} \nabla^{(\eta)}_{\tau} \chi^{\nu} +  \eta^{ \nu \tau} \nabla^{(\eta)}_{\tau} \chi^{\mu}.
\end{equation}
The group of Weyl transformations and transverse diffeomorphisms acting on the linear theory was argued in~\cite{Alvarez2016b} to have the structure of a semidirect product $G = \textrm{LWeyl} \rtimes \textrm{LTDiff}$. Now we will explore possible non-linear completions of both theories and compare them. 

\subsection{Einstein-Hilbert action: Diff or WTDiff principle}
\label{Subsec:Einstein_Hilbert}

In this section we will work in arbitrary spacetime dimension $D+1$, since the generalization is straightforward and will be needed later. Both of the theories from Subsec.~\ref{Subsec:Classica_Linear_Theory} correspond to the linearized version of GR and the WTDiff-invariant UG version of GR, respectively.

GR can be obtained from a variational principle in which the action is diffeomorphism invariant. Diffeomorphism invariance dictates that the unique volume form that we can use to integrate is the one associated with the metric tensor, i.e. $ \frac{1}{(D+1)!} \sqrt{|g|} dx^0 \wedge ... \wedge dx^D$. In a local chart, this reduces to an integration with respect to the weight $d^{D+1}x \sqrt{|g|}$ up to a normalization constant. Now, we need to choose local diffeomorphism invariant objects to build an action. In principle, any scalar of curvature would serve for this purpose. The Ricci scalar $R(\boldsymbol g)$ with a possible cosmological constant term is the lowest dimensional operator that we can add to the action and leads to second order equations of motion for the metric in arbitrary dimensions. It is called Einstein-Hilbert action and it is given by the following expression~\cite{Wald1984}
\begin{equation}
S_{\text{EH}}[\boldsymbol{g}] = \frac{1}{2 \kappa^2} \int_{\mathcal{M}} d^{D+1} x \sqrt{|g|} \left[-2 \Lambda + R (\boldsymbol{g}) \right],
\label{einstein_hilbert_diff}
\end{equation}
where $\boldsymbol{g}$ denotes the metric of the manifold and $\Lambda$ is the cosmological constant. We can add another piece to the action representing additional matter fields, $S_{\textrm{matter}}[\Phi,\boldsymbol{g}]$, where $\Phi$ collectively represents all of the matter fields present in the theory. Diffeomorphism invariance means that the action is invariant under the following transformations generated by a vector field $\xi^{\mu}$ (we lower and raise indices with the metric $g_{\mu \nu}$ and its inverse $g^{\mu \nu}$, respectively)
\begin{equation}
    g_{\mu \nu} \rightarrow g_{\mu \nu} + \delta_{\xi} g_{\mu \nu}, \qquad \delta_{\xi} g_{\mu \nu} = 2 \nabla_{( \mu} \xi_{\nu)}. 
    \label{gdiffstransform}
\end{equation}
Looking for stationary configurations of the action and dropping boundary terms, we find the Einstein equations as the equations of motion of the gravitational field $g_{\mu \nu}$
\begin{equation}
R_{\mu \nu} (\boldsymbol{g})- \frac{1}{2}  g_{\mu \nu} R(\boldsymbol{g}) + \Lambda g_{\mu \nu} = \kappa^2 T_{\mu \nu},
\end{equation}
where 
\begin{align}
T_{\mu \nu} \left( \boldsymbol{g} \right) = \frac{-2}{\sqrt{|g|}} \frac{\delta S_{\textrm{matter}}}{\delta g^{\mu \nu}}
\label{Eq:EM_Tensor_Definition}
\end{align}
is the energy-momentum tensor. 

Let us now write down a theory in which the guiding principle is not that of a Diff invariance but that of WTDiff invariance. First, we note that we need to introduce a privileged background volume form~\cite{Carballorubio2015} which we represent as $\boldsymbol{\omega} = \frac{1}{(D+1)!} \omega(x) dx^0 \wedge ... \wedge dx^D$. In our considerations here, this volume form will always be non-dynamical. For the purpose of building an action, it simply means that in a local coordinate chart we will have a measure of integration in the manifold of the form $d^{D+1}\, x \omega(x) $. With this volume form, we can build the following Weyl-invariant auxiliary metric
\begin{equation}
    \tilde{g}_{\mu \nu} = g_{\mu \nu} \left( \frac{\omega^2}{|g|} \right)^{\frac{1}{D+1}}.
    \label{auxiliary_metric}
\end{equation}
Every curvature scalar built from the auxiliary metric $\tilde{g}_{\mu \nu}$ will be automatically invariant under transverse diffeomorphisms and Weyl transformations (since the auxiliary metric is already Weyl invariant). These transformations are the gauge symmetries of this theory. Explicitly, they correspond to   Weyl transformations, generated by a scalar field $\varphi$ and can be represented infinitesimally as
\begin{equation}
    \delta_{\varphi} g_{\mu \nu} =  \frac{1}{2} \varphi g_{\mu \nu}, 
    \label{infinitesimal_Weyl_transformation}
\end{equation}
and transverse diffeomorphisms, those generated by a vector fields $\xi^{\mu}$ that leave invariant the volume form $\boldsymbol{\omega}$ in the sense of having vanishing Lie derivative
\begin{equation}
    \mathcal{L}_{\xi} \boldsymbol \omega = 0 \quad\rightarrow \quad\tilde{\nabla}_{\mu} \xi^{\mu} = 0, 
\end{equation}
where $\tilde{\nabla}$ is the derivative compatible with the auxiliary metric $\tilde{g}_{\mu \nu}$. The transverse diffeomorphisms act on the metric field as 
\begin{equation}
    \delta_{\xi} g_{\mu \nu} = 2 \nabla_{( \mu} \xi_{\nu)},
\end{equation}
with $\xi_{\nu} = g_{\nu \mu} \xi^{\mu}$. The transverse condition, which is the zero-divergence condition on the vector $\xi^{\mu}$ with respect to the covariant derivative compatible with the auxiliary metric $\tilde{g}_{\mu \nu}$, can be equivalently written as  
\begin{equation}
    \tilde{\nabla}_{\mu} \xi^{\mu}  = 0\quad \rightarrow\quad \nabla_{\mu} \xi^{\mu} = - \frac{1}{2} \xi^{\sigma} \partial_{\sigma} \log \left( \frac{\omega^2}{\abs{g}} \right),
    \label{transverse_diff_divergence}
\end{equation}
in terms of the covariant derivative compatible with the dynamical metric $g_{\mu \nu}$. Notice that this can be understood as a consequence of the nature of the covariant derivative $\tilde{\nabla}$, which is the covariant derivative compatible with an integrable Weyl connection~\cite{Salim1996,Romero2012,Yuan2013,Barcelo2017}. Finally, notice that these transformations can be equivalently expressed as transformations acting on the auxiliary metric $\tilde{g}_{\mu \nu}$ as follows: 
\begin{equation}
    \delta_{\xi} \tilde{g}_{\mu \nu} = 2 \tilde{\nabla}_{(\mu} \tilde{\xi}_{\nu)}, \qquad \tilde{\nabla}_{\mu} \xi^{\mu} = 0, 
    \label{variation2}
\end{equation}
where $\tilde{\xi}_{\mu} = \tilde{g}_{\mu \nu} \xi^{\nu}$. Notice also that $\tilde{\xi}_{\mu} = \left( \frac{\omega^2}{\abs{g}} \right)^{\frac{1}{D+1}} \xi_{\mu}$. This concludes our discussion of the gauge symmetries of a WTDiff-invariant theory. 

Furthermore, writing everything in terms of the auxiliary metric ensures that just the conformal structure of spacetime fluctuates, with the volume form of the geometry to which we couple matter fields and particles fixed \emph{a priori}. With this, the lowest order operator that we can write down is the Ricci scalar in terms of $\tilde{\boldsymbol{g}}$. The cosmological constant operator is no longer a dynamical term in the action since it is independent of $\boldsymbol{g}$, and hence it simply corresponds to adding a constant to the action. With this, we can now write down the action
\begin{equation}
    \tilde{S}_{\text{EH}}[\boldsymbol{g}; \boldsymbol{\omega}] = S_{\text{EH}}[\boldsymbol{\tilde{g}}] = \frac{1}{2 \kappa^2} \int d^{D+1} x\,\omega  R ( \boldsymbol{\tilde{g}} ) ,
\label{einstein_hilbert_wtdiff}
\end{equation}
where, although we use the object $\tilde{g}_{\mu \nu}$ as the metric with respect to which we compute the scalar of curvature and we build the matter action, we consider the field $g_{\mu \nu}$ as the dynamical field. Furthermore, the action of matter that we would add now would be the same we had before up to replacing $\boldsymbol{g}$ by the auxiliary metric $\boldsymbol{\tilde{g}}$:  $ \tilde{S}_{\textrm{matter}} (\Phi, \boldsymbol{g}) = S_{\textrm{matter}} (\Phi, \boldsymbol{\tilde{g}})$. 

This formulation ensures that the variation with respect to $g_{\mu \nu}$ always leads to traceless equations. This can be seen from the relation between $\tilde{g}_{\mu \nu}$ and $g_{\mu \nu}$ in Eq.~\eqref{auxiliary_metric}, since
\begin{equation}
    \delta \tilde{g}^{\mu \nu} = \delta g^{\mu \nu} - \frac{1}{D+1} g^{\mu \nu} g_{\alpha \beta} \delta g^{\alpha \beta}.
\end{equation}
Notice that $g_{\mu \nu} g^{\alpha \beta} = \tilde{g}_{\mu \nu} \tilde{g}^{\alpha \beta}$ since the conformal factor that enters multiplying is cancelled by the one that enters in the denominator. With this, the variation of the Ricci-scalar part of the action reads
\begin{equation}
    \delta \tilde{S}_{\text{EH}} [\boldsymbol{g}] = \int d^{D+1} x\, \omega \left(R_{\mu \nu} (\boldsymbol{\tilde{g}}) - \frac{1}{D+1} R (\boldsymbol{\tilde{g}}) \tilde{g}_{\mu \nu } \right) \delta g^{\mu \nu}.
\end{equation}
The variation of the matter action, on the other hand, gives the traceless part of the energy-momentum tensor automatically, since the metric that enters the matter action is $\tilde{g}_{\mu \nu}$. 
\begin{equation}
    R_{\mu \nu} (\boldsymbol{\tilde{g}}) - \frac{1}{D+1} R (\boldsymbol{\tilde{g}}) \tilde{g}_{\mu \nu } = \kappa^2 \left( T_{\mu \nu} (\boldsymbol{\tilde{g}}) - \frac{1}{D+1} T (\boldsymbol{\tilde{g}}) \tilde{g}_{\mu \nu} \right).
    \label{traceless_einstein}
\end{equation}
Taking the covariant derivative, assuming the conservation of the energy-momentum tensor $\tilde{\nabla}_{\mu} T^{\mu \nu} (\boldsymbol{\tilde{g}}) = 0$~\cite{Ellis2010,Ellis2013}, and applying Bianchi identities $\tilde{\nabla}_{\mu} \left[ R^{\mu \nu} (\boldsymbol{\tilde{g}})- R (\boldsymbol{\tilde{g}})  \tilde{g}^{\mu \nu} / 2 \right] = 0$, we can find
\begin{equation}
    \tilde{\nabla}_{\mu} R (\boldsymbol{\tilde{g}}) = -  \kappa^2 \tilde{\nabla}_{\mu} T (\boldsymbol{\tilde{g}}),
\end{equation}
i.e. the Ricci scalar is related to the trace of the energy-momentum tensor
\begin{equation}
R (\boldsymbol{\tilde{g}}) +  \kappa^2 T (\boldsymbol{\tilde{g}}) = 4\Lambda ,
\end{equation}
with $\Lambda$ being an arbitrary integration constant. Plugging it on Eq.~\eqref{traceless_einstein} we find
\begin{equation}
    R_{\mu \nu} (\boldsymbol{\tilde{g}}) - \frac{1}{2} R (\boldsymbol{\tilde{g}}) \tilde{g}_{\mu \nu } + \Lambda \tilde{g}_{\mu \nu}= \kappa^2 T_{\mu \nu} (\boldsymbol{\tilde{g}}),
\end{equation}
which are equivalent to Einstein equations upon performing the gauge fixing $\omega = \sqrt{|g|}$. To understand why this condition is a gauge fixing, we first need to recall that the finite Weyl transformations whose infinitesimal version is that of Eq.~\eqref{infinitesimal_Weyl_transformation} correspond to the following transformations
\begin{equation}
    g_{\mu \nu} \rightarrow e^{\varphi} g_{\mu \nu}.
    \label{finite_weyl}
\end{equation}
From this equation it is clear that we can always perform a Weyl transformation that fixes the determinant of the metric to the desired functional profile. We emphasize that Weyl invariance does not automatically imply conformal invariance. Whereas Weyl invariance refers to the invariance of the action under conformal rescalings of the metric only (no action on matter fields whatsoever), conformal invariance implies also a suitable compensating rescaling of the fields involved in the construction~\cite{Birrell1982}. 

Notice that the cosmological constant here enters as an integration constant, not as a constant in the Lagrangian. This is the main difference from WTDiff-invariant theories and Diff-invariant ones, the implications of which will be discussed extensively later. Up to now, we have shown that for every solution of GR with a given cosmological constant parameter, we have the same solution in UG. The main difference is that now the cosmological constant is an integration constant. 

Although some previous works have reported potential differences among UG and GR at the perturbative level in cosmological scenarios~\cite{Gao2014}, our results point that both theories are completely equivalent at the non-perturbative level. Hence, we think that the potential differences reported should arise as a consequence of the perturbative treatment employed there, and that a more careful analysis of the issue is desirable. 

\subsection{Higher-derivative generalizations}
\label{Subsec:Higher_Derivatives}

In this section we will prove that higher-derivative generalizations of the Einstein-Hilbert action with Diff invariance~\eqref{einstein_hilbert_diff} and  WTDiff invariance~\eqref{einstein_hilbert_wtdiff} are classically equivalent in everything but the cosmological constant, which always reenters the equations of motion as an integration constant. To the authors' knowledge, this point has not been discussed before in the literature in a systematic way. 

\paragraph*{\textbf{Diff-invariant theories.}}

We start with Diff-invariant generalizations of General Relativity: consider a general theory with as many higher curvature terms as one wants, like $f(R)$ theories, for instance. The only requirement for the action is to be invariant under diffeomorphisms. We will have  
\begin{equation}
    S_{\text{Diff}} [\boldsymbol{g}]= \int d^{D+1}x \sqrt{|g|} \big[f(R^{\alpha}_{\ \beta  \gamma \delta} (\boldsymbol{g}) , \boldsymbol{g})-2\Lambda\big],
\label{action_diff_higher}
\end{equation}
where the function $f$ is a scalar of curvature built from the metric tensor and its derivatives, through the Riemann tensor. The action can also contain derivatives of the curvature tensor, i.e. more precisely we would need to write down $f\left(\boldsymbol{g}, R^{\alpha}_{\ \beta  \gamma \delta} (\boldsymbol{g}), \nabla_{\gamma} R^{\alpha}_{\ \beta  \gamma \delta} (\boldsymbol{g}) \cdots \right) $ although we will skip the derivatives to avoid a cumbersome notation. The variation of such an action will contain three pieces
\begin{equation}
    \delta S_{\text{Diff}} [\boldsymbol{g}] = \int d^{D+1} x \sqrt{|g|} \Big[ E_{\mu \nu} (\boldsymbol{g}) - \frac{1}{2} g_{\mu \nu} f \left( R^{\alpha}_{\ \beta  \gamma \delta} (\boldsymbol{g}) , \boldsymbol{g} \right) +\Lambda g_{\mu\nu}\Big] \delta g^{\mu \nu} ,
\end{equation} 
where the first term corresponds to the variation of the $f(R^{\alpha}_{\ \beta  \gamma \delta} (\boldsymbol{g}) ,\boldsymbol{g})$ piece of the action, while the last two terms correspond to the variation of the determinant of the metric. We will assume that we have added all of the suitable Gibbons-Hawking-York like terms~\cite{York1972,Gibbons1978}, which are needed to ensure having a well-posed variational problem in which all the boundary terms vanish if the variation $\delta g^{\mu\nu}$ vanishes at the boundary of the manifold. Explicitly, the equations of motion of this theory are
\begin{equation}
     E_{\mu \nu} (\boldsymbol{g}) - \frac{1}{2} g_{\mu \nu} f(R^{\alpha}_{\ \beta  \gamma \delta} (\boldsymbol{g}) , \boldsymbol{g} ) +\Lambda g_{\mu \nu} = 0. 
\end{equation}
Now we will use the fact that the theory is invariant under diffeomorphisms. Let us begin with the case $\Lambda = 0$, since of course the action is diffeomorphism invariant independently of the value of $\Lambda$. As such, if we take a variation of the form $\delta g^{\mu \nu} = 2 \nabla^{(\mu} \xi^{\nu)}$ such that it vanishes on the boundary, we will have $\delta S = 0$. This means that we have
\begin{align}
    0 =  &  
    \int d^{D+1} x \sqrt{-g} \Big[ E_{\mu \nu} (\boldsymbol{g}) - \frac{1}{2} g_{\mu \nu} f(R^{\alpha}_{\ \beta  \gamma \delta} (\boldsymbol{g}) , \boldsymbol{g} ) \Big] \nabla^{(\mu} \xi^{\nu)}  \\
    = & - 
    \int d^{D+1} x \sqrt{-g} \xi^{\nu} \nabla^{\mu} \Big[ E_{\mu \nu} (\boldsymbol{g})  - \frac{1}{2} g_{\mu \nu} f(R^{\alpha}_{\ \beta  \gamma \delta} (\boldsymbol{g}) , \boldsymbol{g}) \Big],
\end{align} 
where we have integrated by parts and discarded boundary terms again. The vector fields $\xi^{\mu}$ are unconstrained, and hence this means that the vanishing of the previous expression implies that we have the identity
\begin{align}
    \nabla_{\mu} \Big[ E^{\mu \nu} (\boldsymbol{g}) - \frac{1}{2} g^{\mu \nu} f(R^{\alpha}_{\ \beta  \gamma \delta} (\boldsymbol{g}) , \boldsymbol{g} ) \Big] = 0,
    \label{Eq:Generalized_Bianchi_Id}
\end{align}
which is a generalization of the contracted Bianchi identities for higher-derivative theories. Despite the derivation we have made from a variational principle, notice that Eq.~\eqref{Eq:Generalized_Bianchi_Id} is simply a geometrical identity. We have merely used the variation with respect to transformations generated by the vector fields $\xi^{\mu}$ as a tool to uncover it. The same happens in General Relativity with the identity $\nabla_{\mu} \big[R^{\mu \nu} (\boldsymbol{g}) - g^{\mu \nu} R (\boldsymbol{g}) /2 \big] =  0$ because the Einstein-Hilbert action in Eq.~\eqref{einstein_hilbert_diff} is diffeomorphism invariant. 

Now, we can repeat the same argument but with a non-vanishing cosmological constant $\Lambda \neq 0$. This means that instead of Eq.~\eqref{Eq:Generalized_Bianchi_Id} we would reach
\begin{align}
    \nabla_{\mu} \Big[ E^{\mu \nu} (\boldsymbol{g}) - \frac{1}{2} g^{\mu \nu} f(R^{\alpha}_{\ \beta  \gamma \delta} (\boldsymbol{g}) , \boldsymbol{g} ) + \Lambda g_{\mu \nu} \Big] = 0.
    \label{Eq:Generalized_Bianchi_Id_CC}
\end{align}
If we now use the identity that we have already derived, i.e., Eq.~\eqref{Eq:Generalized_Bianchi_Id}, the expression~\eqref{Eq:Generalized_Bianchi_Id_CC} reduces to the identity $\nabla^{\nu} \Lambda = 0$, which leads to the conclusion that the cosmological constant need to be indeed a constant.

\paragraph*{\textbf{WTDiff-invariant theories.}}

Let us now discuss the same version of the action but with a fixed volume background, i.e., the closest action to~\eqref{action_diff_higher} obeying a WTDiff-invariance principle. For this WTDiff-invariant theory, the action would read
\begin{equation}
    S_{\text{WTDiff}} [\boldsymbol{g}] = S_{\text{Diff}} [\boldsymbol{\tilde{g}}] = \int d^{D+1} x\, \omega f \left[ R^{\alpha}_{\ \beta  \gamma \delta}  ( \boldsymbol{\tilde{g}}), \boldsymbol{\tilde{g}} \right]. 
\label{generic_higher_derivative_wtdiff}
\end{equation}
which is the same function we had for the GR generalization, but instead of writing the curvature scalars terms of the metric $g_{\mu \nu}$, we do it in terms of the metric $\tilde{g}_{\mu \nu}$ metric introduced in Eq.~\eqref{auxiliary_metric}. There is no cosmological constant here because it is non-dynamical, as in the previous section. The equations of motion are now computed as follows
\begin{equation}
    \delta S_{\text{WTDiff}}  [\boldsymbol{g}]  = \int d^{D+1} x\, \omega E_{\mu \nu} ( \boldsymbol{\tilde{g}} ) \delta \tilde{g}^{\mu \nu},
\end{equation}
which can be rewritten in terms of the metric $\boldsymbol{g}$ using
\begin{equation}
\delta \tilde{g}^{\mu \nu} = \delta g^{\mu \nu} - \frac{1}{D+1} g^{\mu \nu} g_{\alpha \beta} \delta g^{\alpha \beta}.
\end{equation}
Again we can use $g_{\mu \nu} g^{\alpha \beta} = \tilde{g}_{\mu \nu} \tilde{g}^{\alpha \beta}$ to express the variation of the action as
\begin{equation}
    \delta S_{\text{WTDiff}} [\boldsymbol{g}] = \int d^{D+1} x\, \omega \left( E_{\mu \nu} (\boldsymbol{\tilde{g}}) - \frac{1}{D+1} \tilde{g}_{\mu \nu}  E (\boldsymbol{\tilde{g}} ) \right) \delta g^{\mu \nu}, 
\end{equation}
where $ E (\boldsymbol{\tilde{g}} ) = \tilde{g}^{\mu \nu} E_{\mu \nu} (\boldsymbol{\tilde{g}}) $. Thus, we have the equations of motion
\begin{equation}
    \tilde{E}_{\mu \nu} ( \boldsymbol{\tilde{g}} ) - \frac{1}{D+1} E ( \boldsymbol{\tilde{g}} ) \tilde{g}_{\mu \nu} = 0. 
\end{equation}
Taking the covariant derivative $\tilde{\nabla}_{\mu}$ and using the geometrical identity~\eqref{Eq:Generalized_Bianchi_Id}, we find that 
\begin{equation}
    \tilde{\nabla}_{\mu} E ( \boldsymbol{\tilde{g}} ) = 0.
\end{equation}
Thus, $E(\boldsymbol{\tilde{g}})$ needs to be a constant. Choosing it adequately to have the same normalization we had for the GR generalization, we reach the following equations of motion
\begin{equation}
    E_{\mu \nu} (\boldsymbol{\tilde{g}}) - \frac{1}{2} \tilde{g}_{\mu \nu} f(R^{\alpha}_{\ \beta  \gamma \delta} (\boldsymbol{\tilde{g}}) , \boldsymbol{\tilde{g}} ) + \Lambda \tilde{g}_{\mu \nu} = 0,
\end{equation}
which are equivalent to the ones obtained in higher-derivative Diff-invariant generalizations of General Relativity upon fixing the gauge $\sqrt{|g|} = \omega$.

\paragraph*{\textbf{Matter fields.}}

The generalization to include an energy-momentum tensor is almost straightforward. If we assume the same matter content is included in both theories and it is described by an energy-momentum tensor $T_{\mu \nu}$, we have the following equations for the Diff-invariant theory
\begin{equation}
   E_{\mu \nu} (\boldsymbol{g})- \frac{1}{2} g_{\mu \nu} f(R^{\alpha}_{\ \beta  \gamma \delta} (\boldsymbol{g}) , \boldsymbol{g} ) +  \Lambda g_{\mu \nu} = \kappa^2 T_{\mu \nu}.
\label{Diff_higher_order_generalizations}
\end{equation}
For WTDiff, we obtain 
\begin{equation}
E_{\mu \nu} ( \boldsymbol{\tilde{g}} ) - \frac{1}{D+1} \tilde{g}_{\mu \nu} E ( \boldsymbol{\tilde{g}})  = \kappa^2 \left( T_{\mu \nu} ( \boldsymbol{\tilde{g}})- \frac{1}{D+1} T ( \boldsymbol{\tilde{g}} ) \tilde{g}_{\mu \nu} \right).
\end{equation}
Taking the derivative and performing the same manipulations introduced above, one reaches the equation 
\begin{equation}
    E_{\mu \nu} (\boldsymbol{\tilde{g}}) - \frac{1}{2} \tilde{g}_{\mu \nu} f(R^{\alpha}_{\ \beta  \gamma \delta} (\boldsymbol{\tilde{g}}) , \boldsymbol{\tilde{g}} ) + \Lambda \tilde{g}_{\mu \nu} = \kappa^2 T_{\mu \nu} (\boldsymbol{\tilde{g}}).
\end{equation}
This set of equations agree with those for a Diff-invariant theory~\eqref{Diff_higher_order_generalizations}, upon fixing the gauge $\sqrt{|g|} = \omega$. Again, $\Lambda$ is an integration constant or a parameter from the action, depending on whether we are on the WTDiff-invariant or the Diff-invariant version. 

As a final comment, let us emphasize that working with a general dynamical connection that is independent of the metric (this means, one does not include the Levi-Civita one), as it is done in the metric-affine formalism, can be easily incorporated in our discussion. For all practical purposes, the general affine connection (typically expressed in terms of the torsion and non-metricity) in those situations would simply be treated as additional matter fields. In that sense, no differences would appear in either approach as long as one does not violate the conservation of the energy-momentum tensor. For instance, this has been checked explicitly in~\cite{Bonder2018,Bonder2019} for Unimodular Einstein-Cartan Gravity. 

\subsection{Reconstruction from the linear theory: Bootstrapping gravity}
\label{Subsec:Bootstrapping}

\paragraph*{\textbf{Diff-invariant theories.}}

The constructibility of GR as a self-interacting theory of graviton fields propagating on top of flat spacetime has been often emphasized as being one of its distinguishing properties. The history of the problem has its roots in Gupta's proposal that a consistent self-interacting theory of gravitons might be unique and it could be General Relativity~\cite{Gupta1954}. The problem was addressed later by many people like Feynman, as it is explained in his lectures on gravitation~\cite{Feynman1996}, where he tried to prove that there was a unique self-consistent theory of gravitons coupling to its own energy-momentum tensor. Although Feynman did not manage to explicitly solve the problem, he argued that the unique solution to such problem needed to have the structure of GR and gave plausible arguments for it. An explicit procedure was later provided by Deser, who showed explicitly by working in first-order formalism that General Relativity was a theory that coupled to its own energy-momentum tensor~\cite{Deser1970}. However, some concerns have been put forward regarding the non-uniqueness of the procedure, see for instance~\cite{Ortin2010,Padmanabhan2008,Barcelo2014}. These concerns are related to the fact that there is an inherent ambiguity in defining a conserved current (in this case the energy-momentum tensor) and hence it gives rise to a potentially infinite ambiguity. 

Let us briefly sketch the idea of the self-coupling of gravitons to their own energy-momentum tensor. The starting point is to consider the theory of linear gravitons, described by a symmetric rank-2 tensor $h^{\mu \nu}$, propagating in flat spacetime as described by Fierz-Pauli theory~\cite{Fierz1939} ($\xi_1=\xi_2=\xi_3=0$ in Eq.~\eqref{masslessspin2}). The idea is to write a source on the right-hand side of the equations of motion built from the field $h^{\mu \nu}$ itself. The free equations of motion are symmetric in their two Lorentz indices and divergenceless. Hence, we need to put on the right-hand side an object that is divergenceless, at least on shell. The natural candidate for such purpose is the energy-momentum tensor of the Fierz-Pauli theory. However, when it comes to derive the right-hand side of the equations of motion from an additional piece of action added to the Fierz-Pauli theory, it involves cubic terms in $h^{\mu \nu}$. Such piece also contributes to the energy-momentum tensor with terms quadratic in $h^{\mu \nu}$. In this way, an infinite series is generated in which one attempts to build an action of the form
\begin{align}
    S = \sum_{n=2}^{\infty} \lambda^{n-2} S_{n}[h^{\mu \nu}; \boldsymbol{\eta}], 
    \label{Eq:All_Order_Action}
\end{align}
where $\lambda$ is introduced as a coupling constant and we keep it as a book-keeping device to determine the number of $h^{\mu \nu}$ each term contains. In other words, each partial action is an expression of the form:
\begin{align}
    S_n = \lambda^n \int d^{D+1} x \sqrt{- \eta } \mathcal{T}^{\mu_1 \mu_2}_{ \ \ \ \ \  \nu_1 \rho_1 \nu_2 \rho_2 \nu_3 \rho_3 \cdots \nu_n \rho_n} \nabla_{\mu_1} h^{\nu_1 \rho_1} \nabla_{\mu_2} h^{\nu_2 \rho_2} h^{\nu_3 \rho_3} \cdots h^{\nu_n \rho_n},
    \label{Eq:Partial_Actions}
\end{align}
where the tensors $\mathcal{T}^{\mu_1 \mu_2}_{ \ \ \ \ \  \nu_1 \rho_1 \nu_2 \rho_2 \nu_3 \rho_3 \cdots \nu_n \rho_n}$ are built from the flat spacetime metric $\eta_{\mu \nu}$, its inverse $\eta^{\mu \nu}$, and the $\delta^{\mu}_{\ \nu}$ tensor. In addition to this, we will have also an expansion of the whole energy-momentum tensor generated by the action~\eqref{Eq:All_Order_Action}, i.e. 
\begin{align}
    & T_{\mu \nu} = \sum_{n=2}^{\infty} \lambda^{n-1} t^{(n)}_{\mu \nu},
\end{align}
where the terms $t^{(n)}_{\mu \nu}$ are computed from the piece of the action $S_n$. We will come back to how they are computed in a moment, when we discuss also the potential ambiguities intrinsic to the energy-momentum tensor. The idea of this bootstrapping is to reconstruct all the orders of the theory by asking the graviton to couple to its own energy-momentum tensor. To put it explicitly, we have the following constraint equations that dictate us how to generate the higher orders of the expansion 
\begin{equation}
    \frac{1}{\sqrt{- \eta}} \frac{\delta S_n}{\delta h^{\mu \nu}} = -  t_{\mu \nu}^{(n-1)},
\end{equation}
with $t_{\mu \nu}^{(n-1)}$ being the energy-momentum tensor obtained from the $n-1$ piece of the action $S_{n-1}$ which needs to be equal to the piece of the equations of motion derived from $S_n$. If one computes it applying Belinfante's procedure~\cite{Rosenfeld1940,Belinfante1940}, one has 
\begin{equation}
    \frac{\delta S_{n-1} [\boldsymbol{\bar{g}}; \boldsymbol{h}]}{\delta \bar{g}^{\mu \nu}} \bigg\rvert_{\boldsymbol{\bar{g}} = \boldsymbol{\eta} } = -  \sqrt{-\eta} t^{(n-1)}_{\mu \nu},
\end{equation}
where $S_{n-1} [\boldsymbol{\bar{g}}; \boldsymbol{h}]$ is the action $S_{n-1}$ where the flat metric $\boldsymbol\eta$ has been replaced by an arbitrary curved metric $\bar{\boldsymbol g}$. Notice that this is a highly non-trivial step, since we are allowed to introduce non-minimal couplings at each order in the bootstrapping procedure that manifest in ambiguities in the energy-momentum tensor. They correspond to identically conserved terms~\cite{Barcelo2014}. This is the source of the potential non-uniqueness of the procedure and, to the authors' knowledge, a proof of the uniqueness of General Relativity is still missing. Further details of the bootstrapping procedure can be found in~\cite{Barcelo2021,Barcelo2021b,Butcher2009}.

The existence of the solution, although already known since Deser's work~\cite{Deser1970}, was explicitly described in the metric formalism in~\cite{Butcher2009}. In that work, Butcher \emph{et al.} performed a reverse engineering exercise in which they showed that every metric theory of gravity obeyed a set of equations which are those of the bootstrapping procedure, namely
\begin{equation}
    \frac{\delta S_{n-1} [\boldsymbol{\bar{g}}; \boldsymbol{h}]}{\delta \bar{g}^{\mu \nu} } = \frac{\delta S_{n} [\boldsymbol{\bar{g}}; \boldsymbol{h}]}{\delta h^{\mu \nu}}.
\label{bootstrapping_fundamental_equation}
\end{equation}
The crucial role of the non-minimal couplings is discussed extensively in that reference, and they explicitly compute the identically conserved pieces that need to be added in order to recover GR through the bootstrapping procedure. 

To put it explicitly, either from Noether's procedure or Hilbert's approach, one can obtain an energy-momentum tensor $T^{\text{C}}_{\mu \nu}$. However, as it is generally the case with conserved currents, it is possible to add an additional identically conserved piece, $T^{\text{I.C.}}_{\mu \nu}$, in such a way that the resulting quantity is still conserved. In other words, the energy-momentum tensor can be written down as 
\begin{align}
    T_{\mu \nu} = T^{\text{C}}_{\mu \nu} + T^{\text{I.C.}}_{\mu \nu} (\alpha_1, \alpha_2, \ldots), 
\end{align}
where $\alpha_1,\alpha_2, \ldots $ are a collection of parameters encoding all the possible ambiguities in the energy-momentum tensor. The clever choice of variables in Deser's original work~\cite{Deser1970} obscured this fact, since for that formalism to work in a finite number of steps one needs to choose the $\alpha_i$-parameters in such a way that the identically conserved piece vanishes $T^{\text{I.C.}}_{\mu \nu} (\alpha_1, \alpha_2, \ldots) = 0$~\cite{Barcelo2014,Carballorubio2019}. 

It should be emphasized that these results hold for arbitrary metric theories, i.e. an arbitrary Riemann-curvature theory [an $f(R^{\alpha}_{\ \beta  \gamma \delta} (\boldsymbol{g}), \boldsymbol{g})$ theory as the ones discussed above]. In~\cite{Deser2017}, Deser argues that the bootstrapping procedure does not work for higher-derivative generalizations of GR in Palatini formalism (treating the metric and the connection as independent variables). However, in Appendix~1 of \cite{Deser2017}, he argues that the bootstrapping procedure works as long as one enforces the connection to be compatible with the metric, i.e., when one restricts to the study of metric theories. It is well known that higher-derivative metric theories are inequivalent in metric and Palatini formalisms, see for example~\cite{Vollick2003,Flanagan2003,Olmo2005}. Both of them just agree for GR. Actually, they typically propagate a different number of degrees of freedom. Hence, we can just expect the Palatini formalism to be equivalent to the metric formalism with respect to the bootstrapping procedure when we impose metric compatibility of the Palatini formalism. Thus, we find that Deser's result from Appendix 1 in~\cite{Deser2017} is compatible with the results of Butcher \emph{et al.}. Furthermore, we expect that the bootstrapping procedure in first-order formalism for higher-derivative theories might require additional information in order to be potentially successful. For instance, it could happen that the ambiguities present in the procedure might allow to complete it. 

\paragraph*{\textbf{WTDiff-invariant theories.}}

If one takes as a starting point the linearized version of WTDiff, we will show that it obeys a set of similar bootstrapping equations. The main difference is that WTDiff couples to the traceless part of its own energy-momentum tensor, as it is to be expected. To show it, one can simply reproduce the computation of Butcher \emph{et al.}~\cite{Butcher2009} considering an action that depends on an arbitrary background structure $\boldsymbol{\Omega}$ and a metric $g^{\mu \nu}$. The arbitrary background structure $\boldsymbol{\Omega}$ could be an additional fixed metric, a fixed background volume form as the $\boldsymbol{\omega}$ appearing in UG theories or any other structure that we put on our manifold. By expanding on top of an arbitrary background metric $\bar{g}^{\mu \nu}$ where we decompose $g^{\mu \nu} = \bar{g}^{\mu \nu} + \lambda h^{\mu \nu}$, it is possible to show that the following identity holds
\begin{equation}
    \frac{\delta S_{n-1} [\boldsymbol{\Omega}; \boldsymbol{\bar{g}}; \boldsymbol{h}]}{\delta \bar{g}^{\mu \nu} } = \frac{\delta S_{n} [\boldsymbol{\Omega}; \boldsymbol{\bar{g}}; \boldsymbol{h}]}{\delta h^{\mu \nu}}, 
\label{bootstrapping_fundamental_equation2}
\end{equation}
where the actions $S_n$ are the terms of order $\lambda^n$ in the action, following the conventions and notation of Butcher \emph{et al.}~\cite{Butcher2009}. We will take the background structure to be the fixed volume form $\boldsymbol\omega$ of WTDiff and apply this result to a generic WTDiff-invariant action of the form~\eqref{generic_higher_derivative_wtdiff}. If we now take into account that the dynamical metric $g_{\mu \nu}$ is always written in terms of the auxiliary metric $\tilde{g}_{\mu \nu}$, as in Eq.~\eqref{auxiliary_metric}, we always have that
\begin{equation}
    \frac{ \delta S_n[\boldsymbol{\omega}; \boldsymbol{\bar{g}}; \boldsymbol{h}]}{\delta g^{\mu \nu}} = \frac{\delta S_n [\boldsymbol{\omega}; \boldsymbol{\bar{g}}; \boldsymbol{h}]}{\delta \bar{g}^{\mu \nu}} - \frac{1}{D+1} \bar{g}_{\mu \nu} \bar{g}^{\alpha \beta } \frac{\delta S_n [\boldsymbol{\omega}; \boldsymbol{\bar{g}}; \boldsymbol{h}]}{\delta \bar{g}^{\alpha \beta}}. 
\end{equation}
By recalling that the energy-momentum tensor is defined as the variation with respect to the auxiliary metric, we immediately find that a theory satisfying a WTDiff principle couples to the traceless part of its own energy-momentum tensor
\begin{equation}
    \frac{\delta S_{n} [\boldsymbol{\omega}; \bar{g}; h]}{\delta h^{\mu \nu}} \bigg\rvert_{\boldsymbol{\bar{g}}= \boldsymbol{\eta}} = -  \sqrt{-\eta} \left( t^{(n-1)}_{\mu \nu}  - \frac{1}{D+1} \eta_{\mu \nu} \eta^{\alpha \beta} t^{(n-1)}_{\alpha \beta} \right)
\end{equation}
With this, we conclude that the result of Butcher \emph{et al.} generalizes to a theory obeying a WTDiff principle, with the caveat that the self-coupling is to the traceless part of its own energy-momentum tensor, instead of the whole energy-momentum tensor. Thus, from the perspective of the self-coupling problem in the metric formalism, both principles 
(Diff and WTDiff) give rise to non-linear theories that self-couple consistently. The question of uniqueness still remains open for both theories. 

\subsection{Non-minimal couplings. Einstein and Jordan frames}
\label{Subsec:Nonminimal_couplings}

Until now, we have been implicitly assuming that all the couplings to gravity are done through a minimal-coupling prescription. Throughout this section, we will explore this potential inequivalence at the classical level of theories with non-minimal couplings based on a Diff and a WTDiff principle respectively. We will first review the well-known results of non-minimal couplings for a Diff-invariant theory, and then we will move on to discuss the case of WTDiff-invariant theories. For the sake of simplicity, we will focus on the Einstein-Hilbert action, i.e., we will just consider the Ricci scalar as a curvature invariant entering the equations of motion. However, our conclusions will hold for arbitrary non-minimal couplings.

\paragraph*{\textbf{Non-minimal couplings in Diff-invariant theories.}}

Non-minimal couplings in Diff-invariant theories can be mapped to an equivalent description in which the theory displays just minimal couplings~\cite{Faraoni1998}. These transformations are associated with conformal rescalings of the metric variable and we refer to this operation as a change of frame. We will first discuss the mathematical grounds on which these transformations are performed and then move on to their physical interpretation. 
 
To begin with, let us consider the following action for a scalar field $\phi$ coupled to the metric $g^{\jordanf}_{\mu \nu}$
\begin{equation}
    S^{\jordanf} [\boldsymbol{g}^{\jordanf},\phi] = \frac{1}{2 \kappa^2} \int d^{D+1} x \sqrt{|g^{\jordanf}|} \left[ F^{\jordanf} (\phi) R (\boldsymbol{g}^{\jordanf}) + G^{\jordanf} (\phi) \partial_{\mu} \phi \partial^{\mu} \phi + V^{\jordanf}  ( \phi ) \right].
    \label{diff_action_jordan}
\end{equation}
where the superscript (J) means that we are working in the so-called Jordan frame. Performing a conformal transformation of the form
\begin{equation}
    g^{\einsteinf}_{\mu \nu} = \left[ F(\phi) \right]^\frac{2}{D-1} g^{\jordanf}_{\mu \nu},
\end{equation}
one finds the action 
\begin{equation}
    S^{\einsteinf} [\boldsymbol{g}^{\einsteinf}, \phi] = \frac{1}{2 \kappa^2} \int d^{D+1} x \sqrt{|g^{\einsteinf}|} \left[ R (\boldsymbol{g}^{\einsteinf}) + G^{\einsteinf} (\phi) \partial_{\mu} \phi \partial^{\mu} \phi + V^{\einsteinf}  ( \phi ) \right],
    \label{diff_action_einstein}
\end{equation}
where the superscript (E) means that we have moved to the so-called Einstein frame~\cite{Faraoni1998}. The functions $G^{\einsteinf} (\phi),V^{\einsteinf} (\phi)$ differ from the ones in the Jordan frame, $G^{\einsteinf} (\phi) \neq G^{\jordanf} (\phi)$ and $V^{\einsteinf} (\phi) \neq V^{\jordanf} (\phi)$. The explicit relation is irrelevant for our discussion and it can be worked out using the standard properties of the Ricci scalar under conformal transformations, see for instance Appendix D of~\cite{Wald1984}. 

There have been many discussions throughout the years about which of the two frames is the ``physical" one, namely, on which of both frames do massive timelike particles propagate on geodesic trajectories~\cite{Faraoni1998}. Here we simply note that the equivalence among both descriptions implies that, if particles propagate in geodesic trajectories in one frame, they will not propagate in geodesic trajectories on a different frame. Putting this into other words, a metric theory of gravity is not specified just by its equation of motion but also by how local probes couple to that geometry~\cite{Sotiriou2007,Will2014}. The same equations of motion with different couplings to local probes would represent different theories. In that sense, one needs to understand that the equivalence among frames requires also the mapping of the probe dynamics from one frame to the other. Hence, specifying a physical theory for gravity coupled to a matter scalar field requires more information than a simple frame: it requires also the dynamics of the probes that will propagate on the corresponding geometry. 

In order to discuss the existence of a theory obeying a WTDiff principle with the same equations of motion in a suitable gauge, let us prove a useful identity which is analogue to~\eqref{Eq:Generalized_Bianchi_Id}. Following the same procedure used above, let us introduce the equations of motion associated with the action from Eq.~\eqref{diff_action_jordan},
\begin{equation}
    K_{\mu \nu} (\boldsymbol{g}^{\jordanf}, \phi) = 0, \quad \textrm{with} \quad K_{\mu \nu} (\boldsymbol{g}^{\jordanf} , \phi) := \frac{\delta S}{\delta g^{\jordanf \mu \nu}} .
    \label{non_minimal_jordan_eqs}
\end{equation}
Using the diffeomorphism invariance of the action under transformations \mbox{$\delta g^{{\jordanf} \mu \nu} = 2 \nabla^{\jordanf( \mu} \xi^{\nu)}$} and performing an integration by parts we find
\begin{equation}
    0 = 2 \int d^{D+1}x \sqrt{|g|} K_{\mu \nu} (\boldsymbol{g}^{\jordanf}, \phi)\nabla^{ \jordanf ( \mu} \xi^{\nu)} = - 2 \int d^{D+1}x\, \xi^{\nu} \nabla^{{\jordanf} \mu} K_{\mu \nu} (\boldsymbol{g}^{\jordanf}, \phi). 
\end{equation}
Thus, we have a generalization of Bianchi identities for a theory with non-minimal couplings
\begin{equation}
    \nabla^{\jordanf}_{\mu} K^{\mu \nu} (\boldsymbol{g}^{\jordanf}, \phi) = 0.
    \label{Bianchi_identity_jordan}
\end{equation}
This relation will be used later.

\paragraph*{\textbf{Non-minimal couplings in WTDiff-invariant theories.}}

For WTDiff theories, we cannot perform frame changes since they are gauge transformations. However, as we have discussed above, frame changes by themselves are not enough to specify a theory, since one needs to specify the frame in which particles propagate in timelike geodesics. In the WTDiff version of these theories, this choice of frame manifests simply as a choice of action. Hence, no superscript associated with frames will be used for these considerations.

First of all, notice that for a generic Diff-invariant theory in the Einstein frame~\eqref{diff_action_einstein}, equivalent to the ones introduced in~\eqref{einstein_hilbert_diff}, we have an equivalent WTDiff-invariant theory, from the family in Eq.~\eqref{einstein_hilbert_diff}. 

The equations of motion of a given Diff-invariant theory in the Jordan frame, such as the one in Eq.~\eqref{diff_action_jordan} with possibly an additional cosmological constant term, can be obtained from a suitable WTDiff-invariant action. For~\eqref{diff_action_jordan}, such theory is given by the following action
\begin{equation}
    \tilde{S}^{\jordanf} [\boldsymbol{g},\phi] = \frac{1}{2 \kappa^2} \int d^{D+1} x\, \omega \left[ F^{\jordanf} (\phi) R (\boldsymbol{\tilde{g}}) + G^{\jordanf} (\phi) \partial_{\mu} \phi \partial^{\mu} \phi + V^{\jordanf}  ( \phi ) \right].
    \label{wtdiff_action_jordan} 
\end{equation}
The variation of this action gives the following equations of motion
\begin{equation}
    K_{\mu \nu} (\boldsymbol{\tilde{g}},\phi) - \frac{1}{D+1} \tilde{g}_{\mu \nu} K ( \boldsymbol{\tilde{g}},\phi )  = 0,
\end{equation}
where $K_{\mu \nu}( \boldsymbol{\tilde{g}},\phi)$ has the same functional dependence on $\boldsymbol{\tilde{g}}$ and its derivatives that $K_{\mu \nu}( \boldsymbol{g},\phi)$ has on $\boldsymbol{g}$ and we have introduced its trace $K ( \boldsymbol{\tilde{g}},\phi) = \tilde{g}^{\mu \nu} K_{\mu \nu} (\boldsymbol{\tilde{g}},\phi)$. Taking the covariant derivative $\tilde{\nabla}_{\mu}$ and using the identity~\eqref{Bianchi_identity_jordan}, we find the following equations upon performing a trivial integration of the resulting equation
\begin{equation}
    K_{\mu \nu} ( \boldsymbol{\tilde{g}},\phi)+ \Lambda \tilde{g}_{\mu \nu} =0. 
\end{equation}
Fixing the gauge in which $\tilde{g}_{\mu \nu} = g_{\mu \nu}$, we reach the equations we had for the Diff-invariant theory, those from Eq.~\eqref{non_minimal_jordan_eqs}. Hence, we have proved that for a generic theory with non-minimal couplings in a Diff-invariant theory, with probe particles following geodesics on either the Jordan frame or the Einstein frame (which correspond to different physical theories), we have equivalent descriptions in a theory obeying a WTDiff-invariance principle. The same statement holds for other kind of theories with other forms of non-minimal couplings.

\subsection{Non-conserved energy-momentum tensor?}
\label{Subsec:Non_Conserv}

We have been working up to now with the assumption that the energy-momentum tensor of the matter sector is a conserved quantity. However, unlike in GR, in UG the conservation of the energy-momentum tensor is not required for consistency of the equations of motion. Let us begin analyzing the GR case\footnote{Notice that the following arguments can be straightforwardly extended to consider higher derivative generalizations of GR.}. Take Einstein equations with an arbitrary energy-momentum tensor. 
\begin{align}
    R^{\mu \nu} - \frac{1}{2} g^{\mu \nu} R = \kappa^2 T^{\mu \nu}.
\end{align}
We can take the divergence now and the left-hand side vanishes even off shell due to the Bianchi identities. Hence, consistency of these equations requires that (at least) on shell, the energy-momentum tensor is divergenceless also, i.e. conserved. For many matter contents, for instance any matter content derived from a diffeomorphism invariant action principle, the conservation of the energy-momentum tensor is ensured from the equations of motion. Let us prove it following~\cite{Ortin2010}. Take an action for the matter sector of the theory, i.e. an action $S_{\text{matter}} \left[ \boldsymbol{g}, \Phi \right]$. If we perform a diffeomorphism transformation on shell (assuming that the variation of the action with respect to the fields vanishes), i.e. a transformation acting only on the metric as $\delta g^{\mu \nu} = 2 \nabla^{(\mu} \xi^{\nu)}$, and such that it vanishes in the boundary
\begin{align}
    0  & = 2 \int d^{D+1} x \sqrt{-g} \frac{\delta S_{\text{matter}}}{\delta g^{\mu \nu}} \nabla^{(\mu} \xi^{\nu)} \\
    & = - \int d^{D+1} x \sqrt{-g} \xi^{\nu} \nabla^{\mu} \left[ \frac{1}{\sqrt{-g}} \frac{\delta S_{\text{matter}}}{\delta g^{\mu \nu}} \right],
\end{align}
where we have passed from the first line to the second line through an integration by parts and discarding boundary terms since the variation vanishes there. If we introduce now the definition of the energy-momentum tensor from Eq.~\eqref{Eq:EM_Tensor_Definition}, we are led to the conclusion that it has to be conserved on-shell $\nabla_{\mu} T^{\mu \nu} = 0$. 

We can illustrate this with an explicit example which is the energy-momentum tensor of a free real scalar field supplemented with the Klein-Gordon equation of motion. The energy-momentum tensor is
\begin{align}
    T_{\mu \nu} = \nabla_{\mu} \phi \nabla_{\nu} \phi - \frac{1}{2}g_{\mu \nu} \nabla_{\alpha} \phi \nabla^{\alpha} \phi, 
\end{align}
we can take the divergence to find after a cancellation among some of the terms that appear:
\begin{align}
    \nabla^{\mu} T_{\mu \nu} =  \nabla^{\mu} \nabla_{\mu} \phi \nabla_{\nu} \phi,
\end{align}
which is automatically zero if we restrict ourselves to the space of solutions, i.e. we go on-shell since $\nabla_{\mu} \nabla^{\mu} \phi = 0$. 

However, there are examples of energy-momentum tensors that are not derived from an action principle, or at least it is not easy to do it. For instance, the energy-momentum tensor of a perfect fluid is given by
\begin{align}
    T^{\mu \nu} = \left( \rho + p \right)u^{\mu} u^{\nu} + p g^{\mu \nu}.
\end{align}
In principle, the conservation is not ensured a priori. Furthermore, the equations of motion of the fluid on the other hand are unspecified. What is usually done is to derive the equations of motion of the fluid by imposing the conservation of the energy-momentum tensor. This leads to the following equations of motion:
\begin{align}
    & \left( \nabla_{\mu} \rho \right) u^{\mu} + \left( \rho + p \right) \nabla_{\mu} u^{\mu} = 0, \\
    & \left( \rho + p \right) u^{\nu} \nabla_{\nu} u^{\mu} + \left( g^{\mu \nu} + u^{ \mu} u^{\nu} \right) \nabla_{\nu} p = 0,
\end{align}
with $p$ the pressure, $\rho$ the density, and $u^{\mu}$ the velocity of the fluid. Once we are within the UG framework, demanding conservation of the energy-momentum tensor is not mandatory anymore, since we can simply allow for its non-conservation. The main problem is that, in general situations, it is not easy to find well-motivated non-conserved energy-momentum tensors. An example where such a non-conservation is motivated from a microscopic principle is~\cite{Perez2017}. In that paper, using some heuristic arguments the authors propose a concrete functional form for an effective-field theory-like expansion of the particular non-conserved energy-momentum tensor divergence current $J_{\nu} = \nabla^{\mu} T_{\mu \nu}$. They make the key assumption that the one form $J_{\nu}$ is exact, i.e. $\nabla_{[\mu} J_{\nu]} = 0$. In this way, it is (at least locally) an exact form that can be written as $J_{\mu} = \nabla_{\mu} Q$. Then, they are able to obtain some effective equations with a cosmological constant where the cosmological constant is the sum of an arbitrary constant and an integral of the current $J_{\mu}$. 

To the authors' knowledge, there is no systematic investigation of the phenomenology of different kind of non-conserved energy-momentum tensors and in the absence of the key assumption that the $J$ form is exact, i.e. there is no way to perform further analysis in general.

\section{Semiclassical regime}
\label{Sec:Semiclassical}

Let us begin the discussion explaining what is understood by semiclassical gravity. Up to now, we have just considered purely classical systems, i.e., we started with an action functional for which the stationary condition leads to the equations of motion of the theory. In that sense, we considered both the gravitational degrees of freedom, which we encode in the metric field $g_{\mu \nu}$ in the discussion above, and the matter fields as classical fields. 

As a step forward towards a theory of quantum gravity, we can consider quantum field theory on curved spacetimes. This means analyzing the properties of the quantum fields once we deviate from Minkowski spacetime and we allow them to propagate on an arbitrary background whose dynamics is frozen. Such a background must be fixed by the classical dynamics where the classical energy-momentum tensor acts as a source for the curvature of spacetime. 

As usual in quantum theory, if these fields are weakly coupled, they will be allowed to fluctuate at small scales. If the fields are not weakly coupled, we consider that we are working with an effective field theory that may involve different degrees of freedom. We expect that a realistic description of this situation would necessarily include these fluctuations as additional contributions to the energy-momentum tensor. In fact, we expect that in this semiclassical limit we can perturbatively account for the backreaction of the quantum fields on the geometry. This would constitute one further step in the results expected from a full theory of quantum gravity. Independently of the ultraviolet completion that we consider for quantum gravity, there has to be a sector of semiclassical states that are well approximated by a smooth spacetime, although the dynamics of such semiclassical states can be affected by quantum corrections. In particular, quantum fluctuations will also contribute to the vacuum energy, giving a contribution to the bare cosmological constant. 

A useful tool to understand the philosophy of semiclassical gravity is the path-integral quantization and the effective-action formalism. The idea behind this approach is to integrate out the field fluctuations, this means averaging over them and looking at their imprint on the equations of motion for the gravitational degrees of freedom. The starting point is the action for the gravitational theory plus the matter  action. We need to consider the gravitational action as an effective action. Hence we need to include all possible operators compatible with the symmetries. Even if we do not write them from the beginning like for the Einstein-Hilbert action~\eqref{einstein_hilbert_diff}, the infinite tower of operators will be generated as we go to higher orders in perturbation theory (at least we need to introduce them with infinite bare couplings to cancel the divergences arising in loop computations). If we take the Einstein-Hilbert action as a starting point,   terms of the form $R^2$, $R_{\mu \nu} R^{\mu \nu}, \textrm{etc}.$ will necessarily be generated by   quantum fluctuations. More explicitly, the idea is to define the effective action $\Gamma[\boldsymbol{g}]$. In general, it would admit a loop expansion in which $\Gamma[\boldsymbol{g}] = \sum_n \Gamma^{(n)}[\boldsymbol{g}]$, where $\Gamma^{(n)}[\boldsymbol{g}]$ is the contribution from the $n$-th loop order in perturbation theory. This means that we are computing
\begin{equation}
e^{i \Gamma(\boldsymbol{g})} = \int \mathcal{D} \Phi\, e^{i (S_{\textrm{bare}} + S_{\textrm{matter}})},
\label{integration_out_effective_action}
\end{equation}
where $S_{\textrm{bare}}$ represents the bare gravitational action containing the Einstein-Hilbert term, the Riemann-squared terms and all of the higher-derivative terms and $S_{\textrm{matter}}$ represents the action of matter fields. The idea is that, for instance, at the one-loop level the cosmological constant term and Newton's constant appearing in the effective action $\Gamma^{(1)} [\boldsymbol{g}]$ will be shifted by   quantum corrections. This is what we mean by   dressing   the cosmological constant due to quantum corrections, as compared to the bare ones in $\Gamma^{(0)} [\boldsymbol{g}]$.

Actually, the one-loop   computation just takes into account the fluctuations of the free fields and it reduces to the computation of certain functional determinants. There are several ways of doing such a computation which apply different regularization schemes for the path integral. All of them have something in common: the process of regularizing the integrals faces the necessity of introducing an arbitrary energy scale $\mu$. It is a standard issue arising in quantum field theory computations and it is at the heart of the renormalization idea. The values of the bare parameters we add to the action are ``meaningless'' in the sense that they are not directly measurable and they have to be renormalized to the values we measure for them. Schematically, we obtain the following for the one-loop action
\begin{equation}
    \Gamma^{(1)}_{\textrm{eff}} (\boldsymbol{g})  = \int d^4x \sqrt{|g|} \big[ - \lambda_0 + \lambda_1 R (\boldsymbol{g})+ \lambda_2 R^2(\boldsymbol{g}) + \lambda_3 C_{\mu \nu \rho \sigma} (\boldsymbol{g})C^{\mu \nu \rho \sigma} (\boldsymbol{g}) \big],
    \label{sakharov_effective_action}
\end{equation}
where we have introduced the Weyl tensor $C^{\mu \nu \rho \sigma} (\boldsymbol{g})$ in the previous expression following the conventions of~\cite{Misner1974} and the coupling constants $\lambda_i$ are generated at one loop. These constants can be separated in three pieces: one depending on the cutoff $\mu$ that we need to introduce to regularize the integral; another one independent of the cutoff, which is finite as we take $ \mu \rightarrow \infty$, generated by the quantum corrections; and another part coming from the bare action $S_{\textrm{bare}} (\boldsymbol{g})$;
\begin{equation}
    \lambda_i = \lambda_i ^{\textrm{bare}}+ \lambda_i ^{\textrm{div}} + \lambda_i ^{\textrm{finite}}.
    \label{one_loop_coupling_constants}
\end{equation}
In this way, now one needs to relate the couplings $\lambda_i$ with the observable quantities and fix them. We will not do it explicitly, since it is irrelevant for our purposes and the decomposition from Eq.~\eqref{one_loop_coupling_constants} depends on the renormalization scheme. 

Consider that instead of the fields propagating on top of a geometry, we assume that they propagate on top of a conformal structure like the one that a generic WTDiff-invariant theory displays. Hence, we would replace everywhere in the action the metric $\boldsymbol{g}$ by the auxiliary metric $\boldsymbol{\tilde{g}}$ from Eq.~\eqref{auxiliary_metric}. Furthermore, we would replace the metric volume form $\sqrt{|g|}$ with the privileged background volume form $\boldsymbol{\omega}$. With this, all the results obtained above would be obtained here but with $\boldsymbol{g}$ replaced by $\boldsymbol{\tilde{g}}$, except for the cosmological constant term, that disappears~\cite{Carballorubio2015}. We will explain the relevance of this observation in detail in the following section. To put it explicitly, we get that for a WTDiff-invariant theory 
\begin{equation}
    \tilde{\Gamma}^{(1)}_{\textrm{eff}} (\boldsymbol{\tilde{g}})  = \int d^4x \,\omega \left( \lambda_1 R (\boldsymbol{\tilde{g}}) + \lambda_2 \tilde{R}^2(\boldsymbol{\tilde{g}}) + \lambda_3 C_{\mu \nu \rho \sigma}(\boldsymbol{\tilde{g}}) C^{\mu \nu \rho \sigma} (\boldsymbol{\tilde{g}}) \right).
    \label{one_loop_effective_action_wtdiff}
\end{equation}
From the perspective of semiclassical gravity, it seems that the difference between a Diff and a WTDiff principle is again restricted to the different behaviour of the cosmological constant.

\subsection{The cosmological constant in WTDiff gravity}
\label{Subsec:CC}

The cosmological constant is one of the problems in theoretical physics to which a lot of attention has been paid~\cite{Weinberg1988,Martin2012,Burgess2013,Padilla2015}. It is sometimes phrased as the huge discrepancy between the measured value of the cosmological constant and its value computed as the vacuum energy of the quantum fields of the Standard Model. However, why the cosmological constant takes a concrete value is something that lies beyond the scope of the effective field theory framework~\cite{Polchinski1992,Manohar2018}. One cannot predict the value of the couplings of the effective field theory and one simply fixes them to their measured value. 

The cosmological constant is the term that enters the effective Einstein equations linearly in $g_{\mu \nu}$. In GR, this term is associated with the coefficient of the operator $\sqrt{-g}$ in the effective gravitational action (in that sense it appears as in GR), whereas in UG it is simply an integration constant that is not related to any of the parameters entering the effective action. 

The problem with the cosmological constant is that it is not technically natural. In simple words, this means that an arbitrary microscopic ultraviolet-complete theory would give rise to a much bigger value for the cosmological constant than the one observed. This is the case for an arbitrary ultraviolet (UV) completion of the Standard Model. To be more precise, let us follow the nomenclature introduced by Burgess~\cite{Burgess2013,Burgess2020}. The cosmological constant is said to be technically natural if one is able to answer the two following questions: i) Why it is small compared to any other scales, ii) Why it remains small when integrating the heavy degrees of freedom to obtain an effective field theory valid to make predictions at the scales of such parameter. Following a Diff-invariant principle, the second question cannot be answered since the cosmological constant operator in Eq.~\eqref{sakharov_effective_action} would take values of the order of the cutoff $\lambda_0 \sim M^4$ with $M$ being the cutoff scale, see e.g.~\cite{Polchinski1992}.

Technical naturalness is usually used as a guiding principle for model building, in a certain sense, as an application of Occam’s razor principle. Explanations based on technically non natural values of some parameters are not easy to accommodate in UV completions of the theory. Hence it is better to find technically natural explanations that are easier to accommodate in UV completions. Although some counterexamples to the naturalness guiding principle are known (see, e.g. \cite{Arkani-Hamed2021}), it has arguably had some success and it is certainly an idea that one might not want to abandon easily. With a Diff-invariant principle the cosmological constant takes an unnatural value. Hence alternative theories in which it takes a natural value are much more desirable for model building.

A specific way of ensuring this is through the presence of an approximate symmetry, i.e. what is called there 't Hooft naturalness as introduced in~\cite{tHooft1979}. However, technical naturalness is a more general concept and there are other ways of ensuring that it is fulfilled~\cite{Burgess2020}. In addition, although the definition of naturalness is introduced by Burgess only for parameters defining the theory, UG makes the cosmological constant entering Einstein equations to be automatically insensitive to the UV completion (and hence it easily accommodates that it takes small values) \cite{Carballorubio2015,Barcelo2018}. Hence we say that in UG the cosmological constant is technically natural in this sense.

Let us explain this point in more detail. The cosmological constant term in Einstein equations enters merely as an integration constant. The question is whether such integration constant is altered by radiative corrections. It was shown in~\cite{Carballorubio2015} that the fixed background volume form $\boldsymbol{\omega}$ protects the cosmological constant from receiving any quantum corrections, hence being completely stable since it is decoupled from the fluctuations of the quantum fields. Apart from this, all the other operators generated by radiative corrections with a Diff principle, like the Riemann-squared terms, appear in WTDiff with the same coefficients, except for the fact that they appear written in terms of the auxiliary metric $\boldsymbol{\tilde{g}}$ from Eq.~\eqref{auxiliary_metric} instead of the dynamical metric $\boldsymbol{g}$. 

There is a point that is worth mentioning here concerning the possibility that the Weyl symmetry becomes anomalous at the quantum level. The well-known conformal anomaly manifests the breaking of the scale invariance of a classical field theory at the quantum level. Naively, one might think that Weyl transformations being local scale transformations, might inevitably be also anomalous. However, one must remember that the conformal anomaly occurs when one tries to quantize a theory in a way that preserves the whole set of diffeomorphism transformations as gauge symmetries. When one insists just on performing a quantization respecting only transverse diffeomorphism invariance, one finds that no conformal anomaly arises, as it was shown in~\cite{Carballorubio2015}.

Let us quickly review the arguments from~\cite{Carballorubio2015}, further details can be found in that reference. The idea is to apply Fujikawa's method~\cite{Fujikawa1979,Fujikawa1980,Weinberg1996} to discern whether a symmetry is anomalous or not. Using the path integral method to quantize a theory requires not only a classical action to quantize but also a path-integral measure which we need to introduce\footnote{As always, to give a proper definition of the path integral one needs to work in Euclidean time and then rotate back to the Lorentzian signature to get sensible results.}, represented by $\mathcal{D} \Phi$ in~\eqref{integration_out_effective_action}. Anomalies are understood in this formalism as situations in which the classical action is invariant under the symmetry but not the measure. The typical procedure to define such measure is to endow the space of fields with an inner product $(\cdot, \cdot)$. For a scalar field, it can be straightforwardly done as follows
\begin{equation}
    (\phi_1, \phi_2) = \int d \mu (x) \phi_1 (x) \phi_2(x),
\end{equation}
where we have introduced the natural measure of integration in a Riemannian space $d \mu (x) = d^{D+1}x \sqrt{|g (x)|}$. In WTDiff, the natural measure would instead be  $d \tilde{\mu} (x) = d^{D+1}x \sqrt{| \tilde{g} (x)|} = d^{D+1}x \omega (x)$, which takes advantage of the privileged background volume form. With this structure of inner product, we can simply consider an orthonormal basis with respect to this product $\{\phi_n \}_n$ such that a generic field in this basis would be decomposed as
\begin{equation}
    \phi (x) =  \sum_n c_n \phi_n(x), \qquad  c_n = \int d \mu (x) \phi_n(x) \phi(x)
\end{equation}
and define the measure of integration that enters the path integral as 
\begin{equation}
    \mathcal{D} \phi = \prod_{n} \frac{d c_n}{\sqrt{2 \pi}}.
\end{equation}
Let us begin with the case of diffeomorphism invariance. In such case, by construction the measure is invariant under both longitudinal and transverse diffeomorphisms. However, it is not invariant under conformal transformations. This can be easily analized in Fujikawa's method by simply computing the Jacobian of the transformation of the measure under conformal transformations. Under an infinitesimal Weyl transformation in which the metric changes as $\delta g_{\mu \nu} = g_{\mu \nu}  \varphi(x)   /2$, the coefficients $c_n$ change as
\begin{equation}
    \delta c_n = \int d \mu (x) \varphi^2(x) \phi_n (x) \phi (x) /4, 
\end{equation}
since the measure changes as
\begin{equation}
    \delta d \mu (x) = d \mu (x) \varphi^2 (x) / 4.
\end{equation}
The Jacobian $J$ associated with this integral is given by 
\begin{equation}
    \log J = \lim_{N \rightarrow \infty} \sum_{n=1}^{\infty} \int d \mu(x) \varphi (x) \phi_n (x) \phi_n(x) / 2,
\end{equation}
where a proper regularization needs to be considered. After such regularization, we find a finite Jacobian which is the manifestation of the conformal anomaly. 

The situation in WTDiff is different because the measure of integration $d \tilde{\mu}(x)$ now is just invariant under transverse diffeomorphisms. This allows it to be invariant under the conformal transformations, since now we have that 
\begin{equation}
    \delta d \tilde{\mu} (x) = d^{4} x\, \delta \omega = 0, 
\end{equation}
because the volume form does not transform under this conformal transformations. Hence, we conclude that there are no Weyl anomalies for scalar fields. The spin $1/2$ and spin $1$ cases can be also straighforwardly worked out and no anomalies are found~\cite{Carballorubio2015}.

Thus, at the semiclassical level, it seems that a WTDiff principle is better than a Diff principle from the perspective of effective field theory: the cosmological constant can naturally take every possible value, even much smaller values than the Planck or electroweak scales. Similar conclusions were reached for perturbative unimodular quantum gravity (a gauge-fixed version of WTDiff) in~\cite{Alvarez2015,Alvarez2015b,Smolin2009}. In~\cite{Smolin2009} it is pointed out that it is just a consequence of the non-anomalous quantization performed, ensuring that the quantum effective action is unimodular given that the original action is also unimodular. The results are extended in~\cite{Smolin2011} to the context of Loop Quantum Gravity. There have been some discussions in the literature, see for instance Weinberg's review on the cosmological constant problem~\cite{Weinberg1988}, on whether a phase transition associated with the spontaneous symmetry breaking of a group at a given energy scale could lead to a jump in the cosmological constant. For the electroweak sector, the situation is even less clear since it seems that there is no phase transition and one simply has a crossover~\cite{Kajantie1996,Kajantie1997}. Assuming conservation of the stress-energy tensor, the behavior of vacuum energy in cosmological phase transitions is the same in GR and UG.

\subsection{Sakharov's induced gravity}
\label{Subsec:Sakharov}

In a seminal work, Sakharov~\cite{Sakharov1967} proposed that the gravitational dynamics could be generated uniquely from the backreaction of the quantum fields propagating on top of a geometry with no dynamics \emph{a priori}. It would correspond to assume that the gravitational field is described as an spectator whose dynamics is induced exclusively by the dynamics of the excitations propagating on top of it. A comprehensive review of this program can be found in~\cite{Visser2002}. 

In few words, the original proposal of Sakharov is that even if the bare coupling constants vanish $\lambda_i^{\textrm{bare}} = 0$, the coupling constants will effectively acquire a finite value if we work with a finite cutoff $\mu$~\cite{Sakharov1967}. After Sakharov's original proposal, several modifications of this idea have been proposed, for instance the proposal of Frolov and Fursaev (see for example~\cite{Frolov1998}) is specially interesting since, although too restrictive for the matter content of the theory, it allows to perform explicit and cutoff-independent computations. In other words, the coupling constants in such models are computable. We refer the reader to the review~\cite{Visser2002} for a discussion of the most relevant proposals. 

If instead of a geometry what we have is a conformal structure, like the one we have in a WTDiff theory, we would find that the same dynamics would be induced by the background structure and Sakharov's proposal is also relevant. Actually, Sakharov's idea is specially interesting for the emergent gravity program. Many physical systems develop regimes in which perturbations can be well described by a collection of relativistic fields propagating on a background geometry or conformal structure~\cite{Barcelo2005}. The dynamics of such fields is not that of a Diff or WTDiff-invariant theory~\cite{Volovik2008,Volovik2009}. However, Sakharov's mechanism provides hints to why this is the case and how one can try to find an emergent theory of gravity~\cite{Barcelo2010,Volovik2008}. Hence, with respect to the induced dynamics \emph{\`a la Sakharov}, Diff and WTDiff principles do not seem to differ in any aspect but the behaviour of the cosmological constant term, again.

\section{Perturbative quantum field theory} 
\label{Sec:Perturbative_QFT}

Until now, we have just regarded the geometry and conformal structure of Diff and WTDiff-invariant theories respectively as classical fields. Throughout this section, we will focus on the perturbative quantization of both of them. This means that we will take the dynamical metric $g^{\mu \nu}$ for Diff and WTDiff theories and break it into a background part $\bar{g}^{\mu \nu}$ and a perturbation representing a graviton field $h^{\mu \nu}$. We focus on the flat spacetime background and we focus on the computation of scattering amplitudes for asymptotic graviton states. Throughout this section, every time we write a metric $g^{\mu \nu}$, it should be understood that it is implicitly written in terms of the background metric $\bar{g}^{\mu \nu}$ and the perturbation $h^{\mu \nu}$.

\subsection{Quantization of linearized Diff and WTDiff theories}
\label{Subsec:Quantization}
Following reference~\cite{deBrito2021}, we will show that there exists a quantization scheme of a Diff and a WTDiff theory in which both of them formally reduce to the computation of the same kind of path integral. Hence, this would prove the perturbative equivalence at the quantum level among both theories. A plausible argument pointing in this direction was given a few years ago in~\cite{Padilla2014}, although it is not a formal conclusive proof. Agreement on the contributions of  Diff-invariant   and unimodular-gravity theories to scattering amplitudes was found in~\cite{Gonzalez-Martin2017,Gonzalez-Martin2018}. Other works have pointed out that both theories seem to be different at the quantum level~\cite{Bufalo2015,Upadhyay2015} or just in the presence of non-minimal couplings~\cite{Herrero-Valea2020}. Our point of view is that since it exists a quantization scheme in which both theories lead to exactly the same results for scattering amplitudes (i.e. leaving the cosmological constant aside), the apparent difference is due to using different schemes in both theories that are not easy to compare.

Let us begin the discussion with an arbitrary Diff-invariant theory. Let us consider a generic diffeomorpism invariant action  $S_{\textrm{Diff}} [\boldsymbol{g}]$  for a metric field $\boldsymbol{g}$. Formally, we are interested in computing the object
\begin{equation}
    Z_{\textrm{Diff}} = \int \frac{\mathcal{D} h_{\mu \nu}}{V_{\textrm{Diff}}} e^{-S_{\textrm{Diff}}[\boldsymbol{g}]},
    \label{perturbative_path_integral}
\end{equation}
where we have introduced the integration measure  $\mathcal{D}h_{\mu \nu}$ which we assume to be Diff-invariant and, hence, no gauge anomaly is present, and we are formally dividing by the (infinite) volume of the gauge group $V_{\textrm{Diff}}$. 

Let us apply Faddeev-Popov procedure~\cite{Faddeev1967} to perform a partial gauge fixing of this theory. The gauge symmetries of this theory are transformations generated by a vector field $\xi^{\mu}$ acting on the metric as in Eq.~\eqref{gdiffstransform}. This vector field $\xi^\mu$ admits a decomposition as a sum of a transverse (divergenceless) part $\xi_\textrm{T}^\mu$ and a longitudinal (irrotational) one that can be written locally as $\nabla^{\mu} \varphi$ 
\begin{align}
    \xi^{\mu} = \xi^{\mu}_{\textrm{T}} + \nabla^{\mu} \varphi, \qquad \nabla_{\mu} \xi^{\mu}_{\textrm{T}} = 0.
    \label{perturbative_gauge_transf_diff}
\end{align}
That every longitudinal diffeomorphism can be written locally as $\nabla^{\mu} \varphi$ is immediate. Global topological obstructions to this might appear~\cite{Nakahara1990} although they are not relevant at the perturbative quantum level. They would show up at a non-perturbative level, similar to Gribov copies~\cite{Gribov1978}.

The subgroup of transverse diffeomorphisms is   generated by transverse vectors $  \xi^{\mu}_{\textrm{T}}$. We want to make a gauge fixing that breaks the Diff group down to the TDiff group, in order to compare with WTDiff, where we will perform a gauge fixing of the WTDiff group down to the TDiff group. Following the standard Faddeev-Popov procedure, we will insert a trivial factor of $1$ in the path integral as follows:
\begin{equation}
    1 = \Delta_{F} (\boldsymbol{g}) \int \mathcal{D} \varphi\, \delta \left[ F \left( \boldsymbol{g}^{\varphi} \right) \right],
    \label{faddeev_popov}
\end{equation}
where $F(\boldsymbol{g}) = 0$ is a particular gauge fixing functional and $\boldsymbol{g}^{\varphi}$ represents the finite transformation associated with a purely longitudinal generator $ \nabla^{\mu} \varphi$, whose infinitesimal action is
\begin{align}
    \delta_\varphi g_{\mu\nu} = 2\nabla_\mu\nabla_\nu\varphi.
\end{align}
Introducing this into the path integral~\eqref{perturbative_path_integral} we find
\begin{equation}
    Z_{\textrm{Diff}} = \int \frac{\mathcal{D} h_{\mu \nu}}{V_{\textrm{Diff}}} \left\{ \Delta_{F} (\boldsymbol{g}) \int \mathcal{D} \varphi\, \delta \left[ F \left( \boldsymbol{g}^{\varphi} \right) \right] \right\} e^{-S_{\textrm{Diff}}[\boldsymbol{g}]}.
\end{equation}
We assume that both the action $S_{\textrm{Diff}}[\boldsymbol{g}]$ and the measure are Diff invariant, and hence, we can perform a change of variables $\boldsymbol{g} \rightarrow \boldsymbol{g}^{\varphi}$. To avoid a cumbersome notation, we also rename $\boldsymbol{g}^{\varphi}$ and write $\boldsymbol{g}$ instead, as usual. Hence the path integral reads 
\begin{equation}
    Z_{\textrm{Diff}} = \int \frac{\mathcal{D} \varphi \mathcal{D} h_{\mu \nu}}{V_{\textrm{Diff}}}  \Delta_{F} (\boldsymbol{g})   \delta \left[ F \left( \boldsymbol{g} \right) \right]e^{-S_{\textrm{Diff}}[\boldsymbol{g}]}.    
    \label{partial_gauge_fix}
\end{equation}
Once we have reached this point, we would like to factor the volume of the Diff group into a piece which is the volume of the TDiff group and another piece. Since the whole set of diffeomorphisms is generated by arbitrary vector fields $\xi^{\mu}$, the $V_{\textrm{Diff}}$ should be written as
\begin{align}
    V_{\textrm{Diff}} = \int \mathcal{D} \xi^{\mu}, 
\end{align}
whereas the volume of the TDiff group should be computed by plugging a $\delta$ functional that ensures that we just integrate over divergenceless vectors
\begin{equation}
    V_{\textrm{TDiff}} = \int \mathcal{D} \xi^{\mu} \delta \left[ \nabla_{\mu} \xi^{\mu} \right]. 
\end{equation}
Both of them can be related~\cite{Percacci2017,Ardon2017} and obey
\begin{equation}
    V_{\textrm{Diff}} =V_{\textrm{TDiff}} \det ( - \nabla^2 )  \int \mathcal{D} \varphi.
\end{equation} 
We immediately see that the integrals over purely longitudinal diffeomorphisms formally cancel in~\eqref{partial_gauge_fix}. Furthermore, to fix the gauge on purely longitudinal diffeomorphisms we can choose the following gauge fixing
\begin{equation}
    F(\boldsymbol{g}) = \abs{g} - \omega^2,
    \label{gauge_fixing1}
\end{equation}
where $\omega^2$ represents a fixed scalar density. The Faddeev-Popov determinant can be computed as 
\begin{equation}
    \Delta_F (\boldsymbol{g}) = \det    (- \nabla^2 ) .
\end{equation}
Collecting everything and plugging it into equation~\eqref{partial_gauge_fix} we reach the following expression for the path integral
\begin{equation}
    Z_{\textrm{Diff}} = \int \frac{\mathcal{D} h_{\mu \nu}}{V_{\textrm{TDiff}}}  \delta (\abs{g} - \omega^2 ) e^{-S_{\textrm{Diff}}[\boldsymbol{g}]}.     
\end{equation}

Let us now perform an analogue gauge fixing for a theory obeying a WTDiff principle. The starting point now is a WTDiff action $S_{\textrm{WTDiff}} [\boldsymbol{g}]$ in which the dynamical metric is $\boldsymbol{g}$. We study formally the WTDiff action which is parallel to the Diff-invariant theory $S_{\textrm{Diff}} [\boldsymbol{g}]$, namely we consider that $S_{\textrm{WTDiff}} [\boldsymbol{g}] = S_{\textrm{Diff}} [\boldsymbol{\tilde{g}}]$, where $\boldsymbol{\tilde{g}}$ is the auxiliary metric introduced in Eq.~\eqref{auxiliary_metric}. Now the path integral reads 
\begin{equation}
    Z_{\textrm{WTDiff}} = \int \frac{\mathcal{D} h_{\mu \nu}}{V_{\textrm{WTDiff}}} e^{-S_{\textrm{WTDiff}}[\boldsymbol{g}]},
    \label{perturbative_path_integral_WTDiff}
\end{equation}
where we have introduced the integration measure   $\mathcal{D}h_{\mu \nu}$ which we assume to be WTDiff invariant, in this case. Instead of $V_{\textrm{Diff}}$, now the factor that appears dividing is the (again infinite) volume of the WTDiff group, namely $V_{\textrm{WTDiff}}$. 

We will now apply also the Faddeev-Popov procedure to fix the gauge symmetry associated with Weyl transformations. The gauge symmetries of the theory are now transformations acting on the metric as 
\begin{align}
    & g_{\mu \nu} \rightarrow g_{\mu \nu} + 2 \nabla_{(\mu} \xi_{\text{T}, \nu)} + \frac{1}{2} \varphi g_{\mu \nu}, \qquad
    \tilde{\nabla}_{\mu} \xi^{\mu}_{\textrm{T}} = 0,
    \label{perturbative_gauge_transf_wtdiff}    
\end{align}
where $\tilde{\nabla}$ is the covariant derivative compatible with the auxiliary metric $\tilde{g}_{\mu \nu}$. 

We now want to perform a partial gauge fixing of the Weyl symmetry in order to prove the formal equivalence between both path integrals. Again, we will do it via Faddeev-Popov procedure by introducing a gauge fixing of the form~\eqref{faddeev_popov}. Let, $\boldsymbol{g}^{\varphi}$ represent the finite transformation associated with a Weyl transformations generated by a scalar field $\varphi$. By the same arguments exposed above, we find
\begin{equation}
    Z_{\textrm{WTDiff}} = \int \frac{\mathcal{D} \varphi \mathcal{D} h_{\mu \nu}}{V_{\textrm{WTDiff}}}  \Delta_{F} (\boldsymbol{g})   \delta \left[ F \left( \boldsymbol{g} \right) \right]e^{-S_{\textrm{WTDiff}}[\boldsymbol{g}]}.    
    \label{partial_gauge_fix2}
\end{equation}
On the one hand, for WTDiff we have that the volume of the gauge group can be factorized as 
\begin{align}
    V_{\textrm{WTDiff}} = V_{\textrm{TDiff}} \times \int \mathcal{D} \varphi,
\end{align}
due to the structure of the gauge group. Furthermore, it is possible to choose the same gauge fixing function as we did before
\begin{equation}
    F(\boldsymbol{g}) = \abs{g} - \omega^2,
\end{equation}
with $\omega$ representing the background volume form. Now the Faddeev-Popov determinant does not contain any differential operator and hence can be absorbed in the measure of integration. Putting everything together, we find the following for the WTDiff path integral
\begin{equation}
    Z_{\textrm{WTDiff}} = \int \frac{\mathcal{D} h_{\mu \nu}}{V_{\textrm{TDiff}}}   \delta ( \abs{g} - \omega^2 )e^{-S_{\textrm{WTDiff}}[\boldsymbol{g}]}.    
\end{equation}
Notice that the auxiliary metric~\eqref{auxiliary_metric} is such that we have the following identity
\begin{equation}
    \delta ( \abs{g} - \omega^2 )e^{-S_{\textrm{WTDiff}}[\boldsymbol{g}]} =  \delta ( \abs{g} - \omega^2 ) e^{-S_{\textrm{Diff}}[\boldsymbol{g}]},  
\end{equation}
since $\delta ( \abs{g} - \omega^2 ) $ ensures that the auxiliary metric $\tilde{g}_{\mu \nu}$ collapses to the dynamical metric $g_{\mu \nu}$, and we have chosen $S_{\textrm{WTDiff}}[\boldsymbol{g}] = S_{\textrm{Diff}}[\boldsymbol{\tilde{g}}]$. Thus, this proves the formal equivalence between both theories at the quantum level when we perform a perturbative quantization. In this way, any scattering amplitude that we compute for which the cosmological constant does not play a role would give the same result in both theories. More formal analysis, for instance the BRST quantization of the theories~\cite{Kugo2021,Kugo2022b,Kugo2022}, are consistent with this picture. 

\subsection{Recursion relations: constructibility of WTDiff-trees}
\label{Subsec:Constructibility}

The proof of the previous section ensures that the perturbative computation of scattering amplitudes of both theories should agree. An explicit computation was performed in~\cite{Alvarez2016} for the four and five point Maximal-Helicity-Violating (MHV) amplitudes. A full agreement with the results for GR was found for these specific amplitudes. In that work, it is mentioned that this is a surprising result, since the propagator and vertices with the gauge fixing chosen there make the intermediate computation very different from those of Diff-invariant GR. Hence, it seems that it is hard to find higher-order scattering amplitudes.  

On-shell methods have proven to be a very useful conceptual tool for unveiling the structure of quantum field theories, see~\cite{Elvang2015} for a review. Specially relevant are the so-called recursion relations for scattering amplitudes, that allow to write $n$-particle scattering amplitudes in terms of the amplitudes involving $n-1$ particles. The first set of these recursive relations proven for quantum field theories were the so-called BCFW (Britto-Cachazzo-Feng-Witten) recursion relations~\cite{Britto2004,Britto2005}, that apply to Yang-Mills theories. A version of such relations for general relativistic scattering amplitudes were later extended to include gravity in~\cite{Benincasa2007,ArkaniHamed2008}. In that sense, it was proven that general-relativistic tree-level scattering amplitudes were constructible according to the definition of~\cite{Benincasa2007b}. See~\cite{Cheung2015} for a more recent discussion. The standard proofs of constructibility require several technical assumptions concerning, for instance, the structure of the propagator in momentum space. As it was already noticed in~\cite{Alvarez2016}, the behaviour of the off-shell propagator in the case of UG does not allow to apply the standard BCFW shift in order to prove the constructibility of on-shell scattering amplitudes for this theory. 

The problem of whether UG tree-level amplitudes are constructible was settled in~\cite{Carballorubio2019}, where it was shown that a BCFW-like recursion relations could be applied to UG. Actually, it was proved that they could be applied to a broader family of theories which included UG.. Instead of using a BCFW shift, in~\cite{Carballorubio2019} it is proved that UG obeys the soft theorems following the idea presented in~\cite{Elvang2016,Laddha2017}, and then, a suitable shift that takes advantage of the structure of the amplitudes constrained to obey the soft theorems, is introduced to prove constructibility. Given that the $n$-particle scattering amplitudes of UG can be expressed in terms of the $n-1$-particle amplitudes and that the  3, 4, and  5-particle scattering amplitudes are identical to those of GR, it is immediate to conclude that the scattering amplitudes at tree level of UG and GR are identical. This is of course consistent with the claims of the previous section.

Recursion relations at an arbitrary number of loops for General Relativity cannot be expected to exist, due to the non-perturbative renormalizability of gravity in vacuum beyond two loops~\cite{tHooft1974}. At the one-loop level, however, pure General Relativity is one-loop finite~\cite{tHooft1974}. Although we still lack a systematic computation of one-loop amplitudes in terms of the tree level amplitudes, it has been possible to express some one-loop scattering amplitudes in terms of trees~\cite{Brandhuber2007}. Again, this is to be expected since the perturbative quantization of both theories performed in the previous section agree. Actually, the computation of the one-loop effective action for WTDiff was performed explicitly in~\cite{Ardon2017,Percacci2017}, and agreement with the computation of the same effective action in a a Diff-invariant theory was found.

\subsection{Asymptotic safety program and WTDiff principle}
\label{Subsec:Asymptotic_Safety}

Asymptotic safety is a program whose aim is to define a theory of quantum gravity as a perturbative quantum field theory. The idea, as originally proposed by Weinberg~\cite{Weinberg1980}, is that the running couplings approach a constant value at high energies. The space of coupling constants is, in principle, infinite. However, it is expected that the critical surface is finite dimensional and hence can be parametrized by a finite number of coupling constants. See~\cite{Percacci2017b, Reuter2019} for comprehensive introductions to the Asymptotic Safety program, and~\cite{Bonanno2020} for a critical review on the current paradigm. See also~\cite{Donoghue2019} for a criticism to the current practice in Asymptotic Safety. Here we will focus on the interplay between such a program and a WTDiff principle. 

The key tool used nowadays in Asymptotic Safety is the so-called Wetterich equation~\cite{Wetterich1993}, that was introduced for gravity in~\cite{Reuter1996}. It is an equation describing the evolution of the effective average action, which we call $\Gamma_k$ that incorporates quantum fluctuations above the scale $k$, as we vary the energy scale $k$ 
\begin{equation}
    k \frac{\partial}{\partial k} \Gamma_{k} = \frac{1}{2} \tr \left[ \left( \frac{1}{\frac{\delta^2 \Gamma_k}{\delta \Phi^i \delta \Phi^i} + R_k } \right) k \frac{\partial}{\partial k} R_k \right],
    \label{Wetterich_equation}
\end{equation}
where $R_k$ is a cutoff function chosen in such a way that suppresses the excitations below the scale $k$~\cite{Percacci2017b} and $\Phi^i$ collectively represents the fields entering the description. We notice that the standard practice in Asymptotic Safety is to work in Euclidean signature. However, it is not clear whether Wick rotation can be done to recover a Lorentzian theory, since nothing ensures that no poles are found when deforming the contour of integration~\cite{Bonanno2020}. However, for our purpose of comparing Diff and WTDiff theories, such issue does not offer any new complication. The first step in the Asymptotic Safety Program is to select a truncation on the basis of operators entering in $\Gamma_k$, containing a finite number of operators $\mathcal{O}_i$, namely 
\begin{equation}
    \Gamma_k = \sum_{i = 1}^M g_i (k) \mathcal{O}_i.
\end{equation}
Then, given an initial set of data at $k = k_0$, Eq.~\eqref{Wetterich_equation} has a well-posed initial-value problem for evolving the coupling constants $g_i(k)$ from $k_0$ to an arbitrary scale $k$. As $k\rightarrow \infty$, one hits a fixed point living on the critical surface. Then, given the ultraviolet fixed point at $k = \infty$, one can obtain the full quantum effective action (within the truncation and to the level approximation computed) as the limit
\begin{equation}
    \Gamma_{\textrm{eff}} = \lim_{k \rightarrow 0} \Gamma_k. 
\end{equation}
The main novelty that offers a WTDiff principle instead of a Diff one in Asymptotic Safety has already been explored in previous sections: the cosmological constant does not receive quantum corrections. Apart from this generic feature of WTDiff theories, several truncations have been studied for UG within the framework of Asymptotic Safety, and no substantial differences have been found with respect to their Diff-invariant versions~\cite{Eichhorn2013,Eichhorn2015} even when matter is included~\cite{deBrito2019,deBrito2020}. 

As it has already been pointed in~\cite{deBrito2021}, quantitative differences might arise due to the application of different quantization schemes, but there exists a quantization scheme that makes them equivalent. We share the point of view of the authors of~\cite{deBrito2021}, specially in virtue of the discussion in previous sections.

\section{Embedding WTDiff in string theory}
\label{Sec:Strings}

It is generally assumed that the low-energy effective field theory containing spacetime fields that describe the massless states of strings is a theory obeying a Diff-invariance principle. Although it is true that one can write down an effective action accounting for the interactions among gravitons  obeying a Diff-invariance principle, it is not true that such effective action is unique. Throughout this section, we will argue that writing the standard general relativistic Diff-invariant action as the effective theory is a decision, and one could equivalently write down a WTDiff-invariant action. There are two arguments that are typically offered in order to argue that the low-energy dynamics of string theory is well captured by a theory obeying a Diff-invariance principle.

The first of these arguments is that it is possible to compute the string scattering amplitudes involving asymptotic graviton states with a low-energy effective description obeying a Diff-invariance principle~\cite{Green1987a,Green1987b}. On the one hand, one computes these scattering amplitudes involving gravitons via correlation functions on the worldsheet in string perturbation theory. Then, one shows that such scattering amplitudes can be computed at low energies equivalently within GR. We will show that the same amplitudes can be computed in UG, obeying a WTDiff-invariance principle. 

The second of these arguments, is that in order to avoid gauge anomalies in the worldsheet when the string propagates on an arbitrary background, one needs to ensure that the so-called $\beta$-functionals vanish~\cite{Polchinski1998a,Polchinski1998b}. They are computed order by order in the string tension $\alpha'$ assuming that the background fields vary much slower than the string scale. Those $\beta$-functionals can be equivalently derived as the equations of motion associated with GR to the lowest-order in the $\alpha'$ expansion, and higher-order Diff-invariant theories at higher orders in the $\alpha'$ expansion, with a suitable matter content associated with the other massless excitations of the theory. If instead of a background geometry $G_{\mu \nu} (X)$ we decide\footnote{This is a decision, since a coherent state of gravitons can be either a geometry or a conformal structure, due to the arguments described in previous sections} that the strings propagate on top of a conformal structure $\tilde{G}_{\mu \nu}$ with a fixed background volume form, a careful study of the $\beta$-functionals would be required. Further discussions on this point will be expanded in a forthcoming publication, in which details on the $\beta$-functional computations for UG will be provided.

\subsection{Graviton scattering from string theory}
\label{Subsec:Graviton_Scattering}

As it is well known, the computation of string $S$-matrix elements reduces to the computation of some on-shell correlation functions with the insertion of certain vertex operators representing the asymptotic states on two-dimensional compact surfaces~\cite{Polchinski1998a,Polchinski1998b,Green1987a,Green1987b}. In particular, we are interested in those asymptotic states to be graviton states. At tree level---the lowest order in the string coupling $g_s$---the dominant topology is that of lowest genus. We will focus our discussion on the tree-level amplitudes. 

Following standard references in string theory~\cite{Polchinski1998a,Polchinski1998b,Green1987a,Green1987b}, to compute the scattering amplitude involving $m$ gravitons with momenta $p_i$ and helicities $h_i$ which we represent as $\mathcal{A}^{(m)} (1^{h_1},2^{h_2},...,m^{h_m})$, one needs to compute a path integral for the Polyakov action $S_P$ that reads schematically
\begin{equation}
    \mathcal{A}^{(m)} (1^{h_1},2^{h_2},...,m^{h_m}) = \frac{1}{g_s^2} \frac{1}{V_\textrm{gauge}} \int \mathcal{D} X \mathcal{D} \boldsymbol{g} \,e^{- S_\text{P} [X,\boldsymbol{g}]} \prod_{i=1}^{m} V_{i} (p_i, h_i),
\end{equation}
where $V_i$ represents the vertex operator associated with a graviton insertion with a given spin and momentum; and $V_\textrm{gauge}$ represents the volume of the gauge symmetry group. Explicitly, the three-point graviton scattering amplitude reads~\cite{Polchinski1998a}
\begin{equation}
    \mathcal{A} (p_1,p_2,p_3; e_1,e_2,e_3) = i \frac{\kappa}{2}(2 \pi)^{26} \delta^{26} \left( p_1 + p_2 + p_3 \right) e_{1 \mu \nu} e_{2 \alpha \beta} e_{3 \gamma \delta} T^{\mu \alpha \gamma} T^{\nu \beta \delta},
\end{equation} 
where
\begin{equation}
    T^{\mu \alpha \gamma} = p_{23}^{\mu} \eta^{\alpha \gamma} + p_{31}^{\alpha} \eta^{ \gamma \mu } + p_{12}^{\gamma} \eta^{\mu \alpha} + \frac{\alpha'}{8} p^{\mu}_{23} p^{\alpha}_{31} p^{\gamma}_{12},
\end{equation}
and $p_{ij}^{\mu} = p^{\mu}_i - p^{\mu}_{j}$. The terms of order $ \order{\alpha'}$ in $T^{\mu \alpha \gamma}$ give $\order{p^4}$ contributions to the scattering amplitude. If we focus just on the terms of order $\order{p^2}$ (which are the relevant ones at low energies), they are equivalent to the ones computed from the terms of order $\order{h^3}$ resulting from the expansion of the Einstein-Hilbert action for the metric $g^{\mu \nu} = \eta^{\mu \nu} + \kappa h^{\mu \nu}$. The same happens with the $m$-graviton amplitudes, if we neglect the higher order contribution from the string amplitude, they agree with those computed from the Einstein-Hilbert action~\cite{Polchinski1998a,Green1987a}. 

As we have discussed in the previous section, general relativistic WTDiff scattering amplitudes at tree level are equivalent to the general-relativistic Diff ones. Hence, this kind of computations do not point towards General Relativity in an unambiguous way. Thus, from the point of view of graviton scattering amplitudes in string theory, WTDiff and Diff principles are equivalent for the low-energy effective field theory.

\section{Non-perturbative path integral formulation: Euclidean quantum gravity} 
\label{Sec:Non_Perturbative_Path_Integral}

Euclidean quantum gravity, conceived as an \emph{ab initio} definition of quantum gravity, is based on defining the path integral for gravity as a sum over Euclidean geometries. This is in close analogy to how we define the path integrals for local quantum field theories in flat spacetime: they are always defined in Euclidean space and the results in Lorentzian signature are obtained after analytic continuation. It is quite natural to argue that a sum over topologies might be needed since we expect topological fluctuations to occur in a theory of quantum gravity~\cite{Wheeler1964,Hawking1979}. Furthermore, Euclidean black-hole saddle-point configurations have   non-trivial topology~\cite{Hartle1976} and their inclusion is required to reproduce the Bekenstein-Hawking formula for black-hole entropy. Hence, the most naive formulation of Euclidean quantum gravity would be based on computing the object
\begin{equation}
    Z = \sum_{\textrm{topologies}} \int \mathcal{D} g \mathcal{D} \phi_{\textrm{matter}} e^{-S_\text{E}[\boldsymbol{g},\Phi]},
    \label{Euclidean_gravity}
\end{equation}
where one needs to sum over inequivalent geometries and topologies. We need to supplement this with suitable boundary conditions which are crucial. For example,  to  compute  the  partition  function  of a  canonical  ensemble  of  spaces  that  asymptote  a  certain  three-geometry,  we  fix  the  time coordinate  of  the  $(3+1)$-decomposition to be a periodic variable with period $\beta$, the inverse of the temperature~\cite{Gibbons1978}. 

However, we must point out that there is no known way of mathematically making sense of the integral in~\eqref{Euclidean_gravity}, except for some two-dimensional models~\cite{Saad2019}. Furthermore, in those models, the path integral seems to give rise to an average over an ensemble of theories instead of a single theory path integral. Thus, strictly speaking, it cannot be an \textit{ab initio} definition of quantum gravity. The best we can do to make sense of the previous expression is to do a perturbative expansion around saddle points of the action that obey the prescribed boundary conditions and add the relevant ones. This means that we can write
\begin{equation}
    Z \approx \sum_{\textrm{saddle points}} e^{-S_\text{E} [\boldsymbol{g}_{\textrm{sad} }, \Phi_{\textrm{sad}}] +  S^{(1)} + \cdots},
\end{equation}
where $S^{(1)}$ represents the one-loop contribution and the ellipsis represent higher-order contributions. From the interpretative point of view it is unclear whether such the sum over all inequivalent Euclidean geometries is relevant for quantum gravity, which is formulated in Lorentzian spacetime. In addition to this problem, we also have the problem of summing over inequivalent topologies. It is well known that four-manifolds are not classifiable and hence making sense of the previous object is hard at best~\cite{Hawking1979}. The WTDiff version of Euclidean gravity does not offer any new insights on these problems of Euclidean quantum gravity. The problem on which it could shed light is the conformal factor problem, that we will review in the next subsection. However, in the last subsection we will show that an analogous problem appears for the Euclidean quantum gravity version of WTDiff, associated with longitudinal diffeomorphisms. Hence, Diff and WTDiff principles can be regarded as being on equal footing from the perspective of Euclidean quantum gravity.

\subsection{The conformal factor problem in the Diff-invariant theory}
\label{Subsec:Conformal_Factor}

There is a severe problem in Euclidean quantum gravity related to the fact that the Euclidean Einstein-Hilbert action is not bounded from below. To illustrate this, we can consider a conformal transformation of the metric appearing in the Einstein-Hilbert action $g_{\mu \nu} \rightarrow \Omega^2 g_{\mu \nu}$ under which it transforms as follows 
\begin{equation}
    S_{\text{EH}}[\boldsymbol{g}]= - \frac{1}{16 \pi \kappa^2} \int d^{4}x \sqrt{g}\, \Omega^2 R (\boldsymbol{g})- \frac{3}{16 \pi \kappa^2} \int d^{4} x \sqrt{g}\, g^{\mu \nu} \nabla_{\mu} \Omega \nabla_{\nu} \Omega. 
\end{equation}
Choosing a conformal factor that oscillates fast with a given frequency, we can make the action arbitrarily large and negative. Consequently, saddle points of the action have negative modes. When negative modes appear, their contribution to the energy is purely imaginary displaying an instability of a classical vacuum against tunneling. Consequently, it would appear that around every classical saddle point there exists an infinite number of negative modes. The main problem is that the gauge symmetry requires to be fixed by a suitable set of constraints to account for the physical degrees of freedom of the theory. One would expect those gauge constraints to remove the negative modes associated to the conformal sector. However, the constraints are imposed in Lorentzian signature and consequently it is not clear how this can be translated to Euclidean space. The interpretation of this potential negative modes around classical saddle points is still an open question obscuring the interpretation of this formulation of quantum gravity (see Section 5.2 of~\cite{Hebecker2018}, for a discussion and a guide to the literature). This problem has been known for a long time and the most common approach to circumvent it is the so called  Gibbons-Hawking-Perry (GHP) prescription. It consists in rotating the contour of integration in such a way that the conformal factor of the metric takes imaginary values~\cite{Gibbons1978}. It might seem that this prescription for dealing with the conformal factor problem is an \emph{ad hoc} imposition which has no physical content and it is an artifact to enforce the convergence of the integral. In~\cite{Gross1982}, the problem of finding the free energy of a gas of free gravitons was solved by path-integral means using the GHP prescription. The expected result is the Stefan-Boltzmann law for particles with two physical polarizations (the two physical polarizations of gravitons) and it is indeed what it was found. Moreover, applying the same prescription to a black-hole background it was also found that it is a bounce which describes the tunneling of hot flat space to nucleate a black hole, displaying the instability of flat space under gravitational perturbations. These two examples, because of being sensible physical results, are evidence supporting the validity of this prescription. However, whether this prescription extends to arbitrary backgrounds or not is still a big open question in Euclidean gravity. A more careful analysis concerning the different way in which one should perform the analytic continuation of the trace and traceless modes was presented in~\cite{Mazur1989} up to one-loop order. The result that follows from the analysis of~\cite{Mazur1989} is that this treatment is equivalent to the GHP prescription, at least for General Relativity at one loop. A further analysis was presented in~\cite{Mottola1995}, where a geometrically-motivated procedure for making sense of the path integral at one loop was developed. Such procedure avoids introducing the standard Faddeev-Popov ghosts that appear when implementing the gauge fixing condition and gives a justification for the GHP prescription.  

\subsection{The longitudinal diffeomorphism problem of WTDiff}
\label{Subsec:Longitudinal_Diff}

For WTDiff theories, the conformal factor problem is not a problem itself since these conformal transformations are gauge transformations.  The potential problem is that longitudinal diffeomorphisms which are not gauge generate also negative modes. Let us consider the Euclidean version of WTDiff in $3+1$ dimensions
\begin{equation}
S_\text{E}[\boldsymbol{g}] = - \frac{1}{2 \kappa^2}\int d^{4} x\, \omega (x) R (\boldsymbol{\tilde{g}}),
\end{equation}
where we have the Ricci scalar constructed out of the reduced metric as 
\begin{equation}
\tilde{g}_{\mu \nu} = g_{\mu \nu} \left( \frac{\omega^2}{g} \right)^{\frac{1}{4}}.
\end{equation}
Let us perform an infinitesimal transformation characterized by a vector field $\xi^{\mu}$ whose action in the auxiliary metric can be written as 
\begin{equation}
    \delta_{\xi} \tilde{g}_{\mu \nu} = 2 \tilde{\nabla}_{(\mu} \tilde{\xi}_{\nu)} + \frac{1}{2} \left[ - \frac{1}{2} \tilde{g}_{\mu \nu} \xi_{\sigma} \partial^{\sigma} \log \left( \frac{\omega^2}{g} \right) - \tilde{g}_{\mu \nu} \nabla_{\alpha} \xi^{\alpha}\right].
\end{equation}
The choice $ \nabla_{\alpha} \xi^{\alpha} = - \frac{1}{2} \xi_{\sigma} \partial^{\sigma} \log \left( \omega^2 / g \right)$ amounts to transverse diffeomorphisms as introduced above in Eq.~\eqref{transverse_diff_divergence}. Had we chosen the divergence of $\nabla_{\alpha} \xi^{\alpha} = - \frac{1}{2} \xi_{\sigma} \partial^{\sigma} \log \left( \omega^2 / g \right) - f$, for some $f$, the variation of the action would be of the form
\begin{equation}
    \delta_{\xi} S \propto \int d^4 x\, \omega f(x) R (\boldsymbol{\tilde{g}}).
\end{equation}
Thus, we reach the same conclusion that we reach for GR. We can choose the mode associated with the transformation of $\xi^{\mu}$ to be as negative as desired, and hence the action is unbounded from below. Thus, the problem of the conformal transformations in General Relativity translates straightforwardly into the longitudinal diffeomorphism problem for WTDiff, as already noticed in the Appendix B of~\cite{Blas2008}. At least in its naive formulation, Euclidean WTDiff does not offer any new insights into the conformal factor mode problem which directly translates into the longitudinal diffeomorphism mode problem. Even if a suitable prescription for dealing with this mode was found, all the other problems of Euclidean gravity would remain. Namely, the problem of the sum over topologies or the difficulty in giving a physical meaning to the results, i.e. rotating to Lorentzian signature to give a physical meaning to the computations remain. 

\section{Conclusions and further work}
\label{Sec:Conclusions}

Let us summarize our results to conclude the paper. Aside from the different behaviour of the cosmological constant in both frameworks, we have not found any other computable difference between Diff or WTDiff theories. The different behaviour of the cosmological constant is intertwined with the existence of a background fiduciary structure (a fixed volume form). In the discussion below we disentangle these two aspects for the sake of exploring in more details their connection and possible avenues for further work.

However, before diving into these aspects, let us briefly stress that there is one potential further difference between Diff-invariant and WTDiff-invariant settings that might arise as a result of violating the assumption of conservation of the energy-momentum tensor, an assumption that we have been making throughout this work. The main problem in this line of research is providing non-conserved energy-momentum tensors that are well motivated. While the relaxation of the conservation principle opens many possibilities, which are difficult to constrain due to the fundamental role that this principle plays, we think that this possibility is certainly worth exploring. However, from a pragmatic point of view one can always add a violation of the energy-momentum tensor by hand in GR, as long as one also adds an additional piece to the equations of motion compensating this non-conservation in order to still fulfil the Bianchi identities. In that sense, models with non-conserved energy-momentum tensors like the one explored in~\cite{Perez2017} can also be embedded in a GR-like framework although it is clear that such modifications of GR are more natural within a UG framework. 

Let us go back to the cosmological constant problem. This problem has been discussed extensively in the literature (see~\cite{Weinberg1988,Martin2012,Burgess2013} and Refs. therein), and is associated with the radiative instability of the cosmological constant in Diff-invariant frameworks. In other words, the cosmological constant is not technically natural in the sense introduced by 't Hooft~\cite{tHooft1979,Burgess2013}. When one relaxes Diff-invariance excluding longitudinal diffeomorphisms, the instability disappears~\cite{Carballorubio2015}. In that sense, whereas with a Diff-invariance principle one finds that the cosmological constant is not technically natural, in a UG theory obeying a WTDiff principle (without longitudinal diffeomorphisms) the cosmological constant is automatically technically natural. Taking into account that no further differences seem to exist between these two frameworks, we think that, from the point of view of model building, a WTDiff principle is superior at this moment if one insists on having a technically natural cosmological constant. Let us stress that the cosmological constant problem is considered to be a crucial problem in modern physics, and the existence of a framework which eliminates this tension is therefore as important as this problem is considered to be. 

The other aspect that characterizes WTDiff is the existence of a privileged background fiduciary structure, in this case a volume form $\boldsymbol{\omega}$. General Relativity does not require any background fiduciary structure although it does not exclude them either, as illustrated by Rosen's reformulation~\cite{rosen1940a,rosen1940b}. We think that modifications of General Relativity that are background dependent are worth exploring, in particular due to our observation in the previous paragraph regarding the cosmological constant problem. Instead of regarding the WTDiff volume form as a sort of frozen degree of freedom that would acquire new dynamics at high energies, we think it is useful to highlight that we can also regard it as a trace of another ``bigger" background structure, like a whole fixed  metric itself. One of the benefits of having a fiduciary background structure, such as a flat spacetime metric for instance, is that it does not allow causal pathologies. In General Relativity, avoiding such pathologies requires the existence of energy conditions and it is not clear which energy conditions are acceptable. Hence, as it has been discussed in~\cite{Barcelo2022}, it is possible to reverse the logic and conjecture that the absence of causal pathologies itself can be the result of having a deeper fixed causal structure. In this way, energy conditions could be a simple consequence of this structure. 

Furthermore, it is of course logically possible to regard this volume form $\boldsymbol\omega$ as a degree of freedom whose dynamics is frozen at low energies, acquiring a certain dynamics at high energies. Although this is of course possible, it might spoil some of the properties of UG theories with respect to the cosmological constant easily. Some models in this direction have been already explored~\cite{Alexander2018,Alexander2020,Alexander2021}, playing with topological field theories in order to keep the cosmological constant protected under radiative corrections. Another suggestive future direction of work is that of considering a bimetric theory of gravity (with an additional metric variable $f_{\mu \nu}$) and some sort of phase transition in which the volume form $\sqrt{-f}$ becomes non-dynamical at low energies. Although such a theory would propagate additional degrees of freedom at high energies (those encoded in the second metric $f_{\mu \nu}$) they would freeze at low energies and its determinant would give rise to a suitable volume form that could be used as the $\omega$ needed in the construction of UG. Additional works which pursue this direction of work in which this volume-form acquires some dynamics are~\cite{Jirousek2018,Hammer2020}.

All of our analysis have been based on the classical and semiclassical regimes, as well as the perturbative quantization of the theory. The non-perturbative quantization is of course something that is not easy to do since, like in GR, there exist many potential ways of facing the problem. We think that one future line of work that may be worth exploring is that of quantizing non-perturbatively (canonical quantization, loop quantization, etc.) some symmetry-reduced models in UG, like minisuperspace models. A pioneer work in this direction was~\cite{Vikman2021}, where the canonical quantization of UG in minisuperspace is used to argue that singularity formation needs to be avoided as a consequence of the uncertainty principle. It has been also suggested~\cite{Herrero-Valea2020} that differences between UG and GR in the presence of non-minimal couplings could still arise when defining gauge-invariant non-linear observables in a non-perturbative regime, which is still an open issue. 

Overall, all evidence points that a WTDiff principle does a better job than a Diff principle in describing current gravitational observations. While this statement can be controversial, it is a trivial corollary of the aspects discussed in this paper. We think this provides strong motivation for the study of WTDiff principle and its possible extensions, in particular regarding the role of background structures.

\begin{acknowledgements}
The authors would like to thank Carlos Barcel\'o for collaboration in early stages of this project and invaluable discussions during the preparation of the manuscript. We would also like to thank Valentin Boyanov, V\'ictor Aldaya, Yuri Bonder, Tom\'as Ort\'in, Roberto Percacci, Oleg Melichev, Alexander Vikman, Alejandro Jim\'enez-Cano, Adri\`a Delhom, Pepe Maldonado Torralba, Jose Beltr\'an Jim\'enez and Mikhail Shaposhnikov for helful conversations. Financial support was provided by the Spanish Government through the projects PID2020-118159GB-C44, and by the Junta de Andaluc\'{\i}a through the project FQM219. GGM acknowledges financial support from the State Agency for Research of the Spanish MCIU through the ``Center of Excellence Severo Ochoa" award to the Instituto de Astrof\'{\i}sica de Andaluc\'{\i}a (SEV-2017-0709). GGM is funded by the Spanish Government fellowship FPU20/01684.
\end{acknowledgements}

\newpage

\appendix 

\section{Alternative formulations of Unimodular Gravity}
\label{App:UG_Formulations}

In this Appendix, we will collect alternative formulations of UG that are used in the literature and discuss the advantages of the approach that we have taken, working with this unconstrained formalism. A reference containing some additional formulations of UG can be found in~\cite{Jirousek2020}. In Subsection~\ref{App:Rep_Inv} we discuss how can we write any theory in a coordinate-invariant way at the expense of making background structures manifest. We illustrate this point with a simple example of a non-relativistic particle. In Subsection~\ref{App:UG_constrained} we discuss the formulaton of UG as a constrained system. In Subsection~\ref{App:HT} we discuss Henneaux and Teitelboim formulation of UG as a constrained system with an additional vector field.
\subsection{Coordinate invariance}
\label{App:Rep_Inv}

It is always possible to make a theory invariant under changes of coordinates or if it is an action involving no spatial dependence, time-reparametrization invariant. For many purposes, it is not useful since it usually involves adding more structure to the equations instead of using the structure at our disposal to simplify them. However, there are two purposes for which it could be useful. First of all, it helps to make background structures in our theory explicit. It is crucial that in order to write an action in a coordinate-invariant way, every background structure needs to be expressed in arbitrary coordinates and, in that sense, it is made explicit. Second, it introduces an additional gauge symmetry, in the sense that it is spurious, that can be used in our benefit. For some calculations it might be much simpler to work in gauges in which the background structure is still manifest. 

Let us illustrate this with an example. Consider the non-relativistic particle that is given by the standard action
\begin{align}
   S [x] = \frac{1}{2} \int dt \left[\left( \frac{dx}{dt}\right)^2-V(x)\right].
   \label{Eq:free_particle}
\end{align}
This action is not invariant under time-reparametrizations, as it can be straightforwardly checked. Under a time reparametrization $t \rightarrow t(\tau) $, we have that the action changes as 
\begin{align}
   S [x]  = \int d \tau \frac{dt}{d \tau} \left[\frac{1}{2} \frac{1}{ ( dt /d \tau ) ^2}  \left( \frac{dx}{d\tau}\right)^2  -V(x)\right].
   \label{Eq:Change_Coords_Newton}
\end{align}
However, we can make it explicitly time-reparametrization invariant by introducing a fixed (non-dynamicaL) one-form $\boldsymbol\omega=\omega(\tau)d\tau$ as 
\begin{align}
    S [x; \boldsymbol\omega] =  \int d \tau \omega(\tau)\left[ \frac{1}{2}\frac{ 1}{\omega(\tau)^2}\left(\frac{dx}{d\tau}\right)^2-V(x) \right]. 
\end{align} 
This action is manifestly invariant under time-reparametrizations since the $\omega(\tau)$-function is introduced in such a way that its transformation law compensates the combination of the transformation law of the integration variable and the function $x(\tau)$. Let us see this explicitly. In Eq.~\eqref{Eq:Change_Coords_Newton}, if we replace the integration element $dt$ by the one-form $\omega(\tau) d\tau$, we automatically make the second term time reparametrization invariant. In the case of the first term, we need to add a factor $\omega^{-2}$ to compensate the change coming from the $dx /d \tau$ piece of the action. In this way, we see that the action with the $\omega$ function is now time-reparametrization invariant. If we choose a time parameter such that $\omega(\tau) = 1$, we find the action~\eqref{Eq:free_particle} we began with. The original action was written down taking the Newtonian absolute time as the time parameter and, hence, the function $\omega(\tau)$ describes such absolute time function in an arbitrary parametrization.

As we advanced, this formulation of the non-relativistic particle makes the background structure (in this case an absolute time function) explicit, at the expense of introducing additional gauge symmetries in our description. Although for the non-relativistic particle it is far from clear how this reformulation could be useful, in Unimodular Gravity it is much easier to work with the background structures manifest, at the expense of enlarging the gauge symmetry of the theory.

\subsection{Unimodular gravity: constrained Einstein-Hilbert action}
\label{App:UG_constrained}

As we explain in detail in the text, Unimodular Gravity is characterized by having a fixed volume form $\boldsymbol{\omega}$ which will be used to fix the determinant of the metric. This means that we want the volume form defined via the metric $\boldsymbol{g}$ to be fixed as
\begin{equation}
    \sqrt{|g|} = \omega,
    \label{fixedvolumeform}
\end{equation}
where $\omega$ is the tensor density associated with the volume form $\boldsymbol{\omega}$. Unimodular Gravity is typically claimed to be General Relativity with the additional constraint $\sqrt{|g|} = 1$. However, this is obviously an identity that can hold just in a particular coordinate system  or a local chart: the left-hand side is a tensor density while the right-hand side is a constant. Hence, it is crucial to notice that, in general, we can plug an arbitrary volume form there.   

Dynamically it is possible to implement this constraint adding a Lagrange multiplier $\lambda$ to the action as
\begin{equation}
S [\boldsymbol{g}] = \frac{1}{2 \kappa^2} \int d^{D+1} x \sqrt{|g|} R(\boldsymbol{g}) + \int d^{D+1} x \lambda \left( \sqrt{|g|} - \omega \right) + S_{\textrm{matter}}(\boldsymbol{g}, \Phi).
\end{equation}
Having fixed the determinant of the metric ensures that we take traceless variations of the metric since $g_{\mu \nu} \delta g^{\mu \nu} = 0$. This means that we end up with the traceless part of Einstein equations
\begin{equation}
R_{\mu \nu} (\boldsymbol{g}) - \frac{1}{D+1}  g_{\mu \nu} R(\boldsymbol{g}) = \kappa^2 \left( T_{\mu \nu} (\boldsymbol{g}) - \frac{1}{D+1}  g_{\mu \nu} T(\boldsymbol{g}) \right). 
\label{tracelesseinstein}
\end{equation}
Using the same arguments used in Section~\ref{Sec:Classical_Theory}, we can end up finding the equivalence of this formulation with the unconstrained formulation introduced in the text. 

\subsection{Unimodular gravity \`a la Henneaux-Teitelboim}
\label{App:HT}

There is another way of defining Unimodular Gravity by introducing an additional vector density $V^{\mu}$. It was originally introduced by Henneaux and Teitelboim~\cite{Henneaux1989}. We must take the following action with $\lambda$ playing the role of a Lagrange multiplier field which will acquire a constant value and will enter Einstein equations as the cosmological constant 
\begin{equation}
S [\boldsymbol{g},\boldsymbol{V}]= \frac{1}{2 \kappa^2} \int d^{D+1} x \sqrt{|g|} R (\boldsymbol{g}) + \int d^4 x \lambda \left( \sqrt{|g|} - \partial_{\mu} V^{\mu} \right) + S_{\textrm{matter}}(\boldsymbol{g}, \Phi).
\label{henneauxteitelboim}
\end{equation}
The equations of motion are the Einstein equations with the role of the cosmological constant being played by $\lambda$ 
\begin{equation}
R_{\mu \nu} (\boldsymbol{g})- \frac{1}{2} g_{\mu \nu} R(\boldsymbol{g}) - \frac{\lambda}{2} g_{\mu \nu} =  \kappa^2 T_{\mu \nu} (\boldsymbol{g}),
\end{equation}
with the constraint
\begin{equation}
\nabla_{\mu} \lambda = 0 \quad\rightarrow\quad \lambda = -2 \Lambda,
\end{equation}
where we have chosen in the last line the constant field to be written in terms of the constant $\Lambda$. It plays the role of the cosmological constant in Einstein equations entering again as an integration constant. Finally, we have the following equation obtained by varying with respect to $\lambda$ 
\begin{equation}
\partial_{\mu} V^{\mu} = \sqrt{|g|}.
\end{equation}
In addition to diffeomorphism invariance, there is a new gauge symmetry associated with the transformations that leave $g_{\mu \nu}$ and $\lambda$ invariant while act on $V^{\mu}$ as 
\begin{equation}
V^{\mu} \rightarrow V^{\mu} + \epsilon^{\mu}, \qquad \textrm{with} \quad \partial_{\mu} \epsilon^{\mu} = 0.
\end{equation}
Within the components of $V^{\mu}$, just $V^0$ has a physical meaning being the other three pure gauge needed to write fully covariant equations. The role played by $V^0$ is that of the zero mode canonically conjugate to the cosmological constant, as we see from~\eqref{henneauxteitelboim}. In this formulation, we manage to have the metric unconstrained at the expense of introducing additional dynamical variables that ensure it. Furthermore, this also ensures that the cosmological constant enters the equations as an integration constant. 

Actually, we can see that the role played by the component $V^0$ is, somehow, the role of the fixed volume form introduced in the previous section. To see this, we just need to notice that if we introduce the following function 
\begin{equation}
T(t) = \int d^Dx V^0,
\end{equation}
we will have that 
\begin{equation}
T(t_2) - T(t_1) = \int_{t_1}^{t_2} dx^0 \int d^D x \partial_{\mu} V^{\mu} = \int_{t_1}^{t_2} dx^0 \int d^D x \sqrt{|g|}.
\end{equation}
It is possible to see that $T(t)$ is a time function conjugate to the cosmological constant (see~\cite{Henneaux1989} for an explicit discussion of this point). Furthermore, it controls the volume element of the spacetime and it is clearly the manifestation in this formalism of having a fixed volume form as a background structure. 

The description presented here, can be also done in terms of a fully antisymmetric $D$-dimensional form and a scalar density taking the dual of the tensorial objects introduced here~\cite{Henneaux1989}.

\bibliography{wtdiff_biblio}

\end{document}